\documentclass[a4paper,fleqn,usenatbib]{mnras}

\usepackage{newtxtext,newtxmath}

\usepackage[T1]{fontenc}
\usepackage{ae,aecompl}

\usepackage{graphicx}	
\usepackage{amsmath}	
\usepackage{amssymb}	
\usepackage{xspace}	

\newcommand {\sixum} {$6$\,$\mu$m\xspace}
\newcommand {\Lsixum} {$L_{\rm 6\mu m}$\xspace}
\newcommand {\Lbol} {$L_{\rm bol}$\xspace}
\newcommand {\lx} {$L_{\rm X}$\xspace}

\newcommand {\ebv} {$E_{B-V}$\xspace}
\newcommand {\nh} {$N_{\mathrm{H}}$\xspace}
\newcommand {\lamedd} {$\lambda_{\mathrm{Edd}}$\xspace}

\newcommand {\ergpersec} {erg~s$^{-1}$\xspace}

\newcommand {\nhunit} {cm$^{-2}$\xspace}

\newcommand{\oiii}{[\ion{O}{iii}]\xspace}

\newcommand{\civ}{\ion{C}{iv}\xspace}

\newcommand {\xmm} {{\it XMM-Newton}\xspace}

\newcommand {\swiftbat} {{\it Swift}~BAT\xspace}

\newcommand {\wise} {{\it WISE}\xspace}

\title[HRQs in X-rays]{X-ray Observations of Luminous Dusty Quasars at $z>2$}

\author[G.~Lansbury et al.]{
G.~B.~Lansbury,$^{1,2}$\thanks{E-mail: george.lansbury@eso.org}
M.~Banerji,$^{1,3}$
A.~C.~Fabian$^{1}$
and M.~J.~Temple$^{1}$
\\
$^{1}$Institute of Astronomy, University of Cambridge, Madingley
  Road, Cambridge, CB3 0HA, UK\\
$^{2}$European Southern Observatory, Karl-Schwarzschild str.\ 2, 85748 Garching bei M\"unchen, Germany\\
$^{3}$Kavli Institute for Cosmology, University of Cambridge,
  Madingley Road, Cambridge CB3 0HA, UK
}

\usepackage{etoolbox}

\makeatletter
\patchcmd\@combinedblfloats{\box\@outputbox}{\unvbox\@outputbox}{}{%
    \errmessage{\noexpand\@combinedblfloats could not be patched}%
}%
\makeatother

\begin{document}
\label{firstpage}
\pagerange{\pageref{firstpage}--\pageref{lastpage}}
\maketitle

\begin{abstract}
We present new X-ray observations of luminous heavily dust-reddened
quasars (HRQs) selected from infrared sky surveys. 
HRQs appear to be a dominant population at high redshifts and the highest
luminosities, and may be associated with a transitional ``blowout'' phase
of black hole and galaxy co-evolution models. Despite this, their
high-energy properties have been poorly known.  
We use the overall sample of $10$ objects with \xmm coverage to study the high-energy properties of HRQs at
$\left< L_{\rm bol} \right> = 10^{47.5}$~\ergpersec and $\left< z \right>= 2.5$. 
For seven sources with strong X-ray detections, we perform
spectral analyses. These find a median X-ray luminosity of
$\left< L_{\rm 2\mbox{-}10\,keV} \right> = 10^{45.1}$~\ergpersec,
comparable to the most powerful X-ray quasars known. 
The gas column densities are $N_{\rm H}=(1$--$8)\times
10^{22}$~\nhunit, in agreement with the amount of dust extinction
observed. The dust to gas ratios are sub-Galactic, but are
higher than found in local AGN. 
The intrinsic X-ray luminosities of HRQs are weak compared to
the mid-infrared (\Lsixum) and bolometric luminosities (\Lbol), in
agreement with findings for other luminous quasar samples. 
For instance, the X-ray to bolometric corrections range from
$\kappa_{\rm bol}\approx 50$--$3000$.
The moderate absorption levels and accretion rates close to the
Eddington limit ($\left< \lambda_{\rm Edd} \right>=1.06$) are in
agreement with a quasar blowout phase. 
Indeed, we find that the HRQs lie in the forbidden region of
the \nh--$\lambda_{\rm Edd}$ plane, and therefore that radiation pressure feedback
on the dusty interstellar medium may be driving a phase of blowout
that has been ongoing for a few $10^{5}$~years. The wider
properties, including \oiii narrow-line region kinematics, broadly agree
with this interpretation. 
\end{abstract}




\section{Introduction}

Growing supermassive black holes (i.e., active
galactic nuclei; AGNs) at the centers of galaxies have an enormous energy
output from accretion, and are thus expected to play a role in the
growth of galaxies. Indeed, tight relations are observed between
properties of black holes and their hosts (e.g.,
\citealt{Magorrian98,Ferrarese00,Gebhardt00}), and theoretical simulations
require AGNs to reproduce the observed properties of massive galaxies
(e.g., \citealt{Springel05,Bower06}). 
A key observational prediction from simulations is that many of the
most rapidly growing supermassive black holes (i.e., luminous quasars)
are enshrouded in gas and dust, and thus obscured at bluer wavelengths
but visible in the infrared through reprocessed dust emission (e.g.,
\citealt{Sanders96,Hopkins08,Narayanan10,Hickox18}). The rapid
accretion should result in a
transitionary reddened quasar phase during which the gas and dust, fuel for stellar and black
hole growth, are being catastrophically disrupted or expelled from the
galaxy. Following this phase of quasar feedback, the
central engine may be more readily observable as an unobscured
quasar. 

If high columns of dust are present around a rapidly accreting
black hole, outflows are expected to be driven by 
radiation pressure (e.g.,
\citealt{Murray05,Fabian08,Debuhr11,Fabian12}), which can strongly
influence the structure and kinematics of circumnuclear and galactic material
(e.g., \citealt{Honig17,Costa18b,Honig19}). 
Evidence of the influence of such processes can be found in low-redshift
X-ray samples, which provide a relatively clean sampling of
the AGN population. In particular, these samples show an absence of objects with high
Eddington ratios ($\lambda_{\rm Edd}=L_{\rm bol}/L_{\rm Edd}\gtrsim
0.1$) {\it and} moderate gas column densities of $\log (N_{\rm H}/\mathrm{cm^{-2}})\approx
22$--$23.5$ (e.g.,
\citealt{Fabian09,Raimundo10,Vasudevan13,Ricci17c,Liu18,Bar19}). 
This is in agreement with models of radiation
pressure feedback on circumnuclear dusty gas (e.g.,
\citealt{Fabian06b,Ishibashi18b}), which predict a 
``forbidden'' or ``blowout'' region in the $\lambda_{\rm Edd}$--$N_{\rm H}$
plane, in which AGN are rarely observed.
The lack of objects discovered in this blowout region to-date may partly result from the
poor sampling of luminous quasars accreting close to the
Eddington limit, and with high dust columns (e.g.,
\citealt{Raimundo10}). \citet{Glikman17b} has already shown that
some luminous red quasars at $z<1$ lie in the forbidden region.
It is crucial to expand such study to the most luminous quasars at their peak
epoch of activity ($z\sim 2$--$3$; e.g., \citealt{Richards06a}).

Following the advent of sensitive large-area optical and infrared surveys of the
sky, a number of luminous reddened quasar samples have been
discovered, which may be fundamentally different from blue quasars
(e.g., \citealt{Klindt19}). In particular, objects with red colours have been identified in the {\it
  WISE} all-sky imaging survey (e.g., \citealt{Eisenhardt12,Hainline14,Tsai15,Assef15}),
in ground-based near-infrared (NIR) imaging surveys (e.g.,
\citealt{Banerji12,Banerji13,Banerji15b,Temple19}), in the SDSS (e.g.,
\citealt{Richards03, Ross15}), and among radio loud objects in the 2MASS
survey (e.g., \citealt{Glikman07,Glikman15}). 
These samples include hyperluminous quasars at $z\gtrsim2$.
At these high redshifts, there is tentative evidence for
an increase in the typical obscuring column around luminous AGNs (e.g.,
\citealt{Aird15a}), and the aforementioned radiative feedback
processes are expected to be common.

A large sample of $z\gtrsim 2$ red quasars with 
spectroscopically identified broad emission lines has been obtained
from a coordinated effort (\citealt{Banerji15b,Temple19}) selecting
and following up red objects from multiple wide-field ($\gtrsim
1000$~deg$^{2}$ in total) NIR surveys: 
the UKIDSS Large Area Survey (ULAS;
\citealt{Lawrence07}); the European Southern Obseratory (ESO)
VISTA Hemisphere Survey (VHS;
\citealt{McMahon13}); and the ESO VISTA Kilo-degree
Infrared Galaxy survey (VIKING; \citealt{Edge13}).
This work has revealed a population of heavily reddened quasars
(``HRQs'' hereafter) 
which have measured extinctions of $0.5 \lesssim E_{B-V} \lesssim 2.0$
(see Section \ref{dust_gas}), and are
therefore typically not detectable in wide-field optical surveys. 
The high dust extinctions and luminosities may correspond
to an enhanced phase of AGN feeding, and possibly feedback. 
Importantly, at the high end of the AGN luminosity function ($L_{\rm bol}\gtrsim 10^{47}$~\ergpersec)
HRQs are at least as numerous as unobscured optical (e.g., SDSS)
quasars (\citealt{Banerji15b}). It is thus of importance to constrain their physical
properties, and better understand this significant phase of black hole growth
in the Universe.

HRQs are compelling laboratories in which to study the accretion
environment around rapidly growing supermassive black holes during
their peak epoch of growth. 
From a practical perspective, the obscuring columns are
sufficiently low that the broad line region is still observable and
the continuum is detected, thus allowing constraints on black hole
mass, bolometric luminosity, and accretion rate. 
From a physical perspective, HRQs appear to be dusty and accreting
at a significant fraction of the Eddington limit (e.g.,
\citealt{Temple19}), meaning that AGN feedback is likely to be having
a significant impact.
Despite this, their high-energy properties are poorly known,
with only three objects having been studied in the X-ray
band (e.g., \citealt{Banerji14,Martocchia17}), and it is unclear how
they fit in with luminous unobscured quasars (e.g., \citealt{Just07,Stern15,Martocchia17}) and other luminous reddened quasars studied at high
energies (e.g., \citealt{LaMassa16c,Glikman17a,Goulding18,Vito18b,Zappacosta18b}).

X-ray photons are emitted from an extremely hot
plasma (the corona) within $\sim 10$ gravitational radii of the
black hole (e.g., \citealt{Fabian15}). Imprinted on the observed X-ray spectrum,
therefore, are the features of absorption and scattering by matter along an
almost direct line to the black hole. It is thus possible to obtain an accurate
measurement of the line-of-sight gaseous column density (\nh) and the intrinsic
(i.e., absorption-corrected)
luminosity. This makes possible: (i) constraints on the makeup and location of the obscuring
media, when comparing to dust extinction; (ii) insight into the physics of
the X-ray corona, when comparing to infrared--ultraviolet emission;
and (iii) tests of AGN feedback models, when comparing
to the accretion rates and outflow properties.

In this paper we report on new X-ray observations which, in combination with
archival data, allow us to characterise the high-energy properties of
the HRQ population, and begin to address
(i)--(iii) above. Section
\ref{sample_data} describes the sample, the observations, and the data
processing and analysis procedures. In Section \ref{results} we
present and discuss the results: measurements of X-ray properties (Section
\ref{xray_props}); dust versus gas obscuration
(Section \ref{dust_gas}); a comparison between the intrinsic X-ray,
infrared, and bolometric accretion luminosities (Section
\ref{lum_section}); and the \nh--$\lambda$ plane for HRQs, in the
context of radiation pressure feedback models (Section
\ref{nhlam_section}). In Section \ref{other_samples} we consider how the
results can be understood in the context of other hyperluminous quasar
samples. Section \ref{summary} summarises the main results. 
We adopt cosmological parameters of $(\Omega_{M}, \Omega_{\Lambda}, h)=(0.27,0.73,0.70)$.

\section{The sample and data}
\label{sample_data}

\subsection{The parent sample}

Our parent sample of highly luminous heavily dust-reddened type~1
quasars (HRQs)
was selected from the large-area ($\gtrsim 1000$~deg$^{2}$) infrared photometric survey
catalogs ULAS (\citealt{Lawrence07}), VHS (\citealt{McMahon13}), and VIKING (\citealt{Edge13}).
The overall approach of the previous studies
(\citealt{Banerji12,Banerji15b,Temple19}) was to
identify objects with extremely red NIR colours, and then perform
subsequent spectroscopic followup to confirm the HRQ candidates as
quasars, and accurately determine redshifts.
The sources considered here have $(J-K)_{\rm Vega}>2.5$ or $(H-K)_{\rm
  Vega}>1.9$, and $K_{\rm Vega}<17.5$.
  They are AGN-dominated
  in the mid-infrared, based on a \wise color selection of
  $(W\mathrm{1}-W\mathrm{2})>0.85$.
  Additionally, six of the sources presented here were also
  required in their initial selections to have optical non-detections,
  faint optical magnitudes, and/or red optical--NIR colors of at
  least $(i-K)_{\rm Vega}>4.4$ (for J1234, J1539, and J2200, refer to Section 2.1 of
  \citealt{Banerji12}; for J2205, J2243, and J2314, refer to Section
  2.1.2 of \citealt{Temple19}).

\subsection{The X-ray sample and observations}

We consider all spectroscopically confirmed HRQs at $z>2$ that have
coverage from sensitive X-ray telescopes. There are ten such quasars,
all with coverage from \xmm, detailed in Table \ref{xrayData_table}. 
For five of these sources (J1122, J2200, J2205, J2243, and J2314), the
quasars were targetted as part of a new \xmm program (PI
Lansbury; program ID 082409), which targetted sources with luminous
\oiii line detections ($L_{\rm [OIII]}>10^{43}$~\ergpersec)
from \citet{Temple19}. J1234 and J2315 were targetted through
our previous programs (PI McMahon and Banerji, respectively), with the
former data published in \citet{Banerji14}. J0144 was covered as part
of the Stripe 82 X-ray survey (\citealt{LaMassa16a}), J1216 was
detected off-axis in an archival observation (PI Ponman), and J1539
was targetted (PI Zappacosta) and published in \citet{Martocchia17}.

The ten X-ray observed HRQs have redshifts in the range
$z=2.085$--$2.658$, and a median redshift of $\left< z\right> =2.503$. 
All objects have broad emission lines in their optical spectra, but are heavily reddened with
$0.53<E_{B-V}<1.94$. 
Sources with $E_{B-V}\gtrsim 1$ are among the dustiest broad-line type 1
quasars known. 
The black hole masses are in the range $8.4< \log M_{\rm BH} < 10.5$, with a median value of 
$9.4$ (\citealt{Temple19}).
The infrared luminosity at \sixum, which traces the intrinsic AGN
power, has a median value of $\left< \log (L_{\rm 6\mu m} / \mathrm{erg\
  s^{-1}})\right> =46.64$. These sources are therefore
some of the most bolometrically luminous quasars.

\setlength{\tabcolsep}{0.3em}  
\begin{table*}
\footnotesize
\centering
\caption{Summary of the \xmm data used for spectral (and photometric) analyses}
\begin{tabular}{llllllllll}
\hline\hline 
Object & R.A. & Decl. & $z$ & Obs ID & UT Date & $t_{\rm on}$ [$\mathrm{ks}$] & $t$ [$\mathrm{ks}$] &
                                                                     $S$ [$\mathrm{cts}$] & $S_{\rm net}/S$ [$\%$] \\
(1) & (2) & (3) & (4)  & (5)  & (6) & (7) & (8) & (9)  & (10) \\
\hline 
\noalign{\smallskip}
ULAS\,J0144$-$0014 & 01:44:52.00 & $-$00:14:32.2 & $2.504$ & 0747440131 & 2014-08-04 & $9.5/11.8$ & $3.5/4.8$ & $26/13$ & $58.0$$/$$63.0$ \\
VHS\,J1122$-$1919 & 11:22:24.43 & $-$19:19:17.4 & $2.464$ & 0824090201 & 2018-07-10 & $15.8/17.6$ & $9.6/16.8$ & $74/49$ & $<$$35.4$$/$$<$$36.9$ \\
ULAS\,J1216$-$0313 & 12:16:31.77 & $-$03:13:35.0 & $2.574$ & 0305800701 & 2005-12-16 & $18.0/26.5$ & $1.8/4.1$ & $36/22$ & $86.0$$/$$87.2$ \\
ULAS\,J1234$+$0907 & 12:34:27.52 & +09:07:54.2 & $2.502$ & 0722960101 & 2013-06-30 & $51.0/52.6$ & $41.8/51.6$ & $528/377$ & $76.4$$/$$81.7$ \\
ULAS\,J1539$+$0557 & 15:39:10.16 & +05:57:49.7 & $2.658$ & 0745010101 & 2015-02-16 & $42.7/44.6$ & $29.5/37.2$ & $261/226$ & $80.7$$/$$84.2$ \\
ULAS\,J2200$+$0056 & 22:00:24.87 & +00:56:04.7 & $2.541$ & 0824090801 & 2018-11-28 & $8.6/13.8$ & $7.3/13.5$ & $210/203$ & $75.9$$/$$84.0$ \\
VIK\,J2205$-$3132 & 22:05:13.68 & $-$31:32:02.4 & $2.307$ & 0824090301 & 2018-11-10 & $42.7/2.8$ & $15.5/\cdots$ & $49/\cdots$ & $<$$36.3$$/$$\cdots$ \\
VIK\,J2243$-$3504 & 22:43:48.96 & $-$35:04:39.4 & $2.085$ & 0824090401 & 2018-11-16 & $22.5/31.5$ & $19.2/30.8$ & $395/336$ & $74.4$$/$$84.4$ \\
VIK\,J2314$-$3459 & 23:14:16.87 & $-$34:59:47.0 & $2.325$ & 0824090501 & 2018-11-14 & $20.3/33.7$ & $17.4/32.9$ & $70/105$ & $53.1$$/$$33.4$ \\
ULAS\,J2315$+$0143 & 23:15:56.23 & +01:43:50.3 & $2.560$ & 0745320101 & 2014-12-16 & $54.0/54.1$ & $28.3/47.4$ & $896/1072$ & $88.3$$/$$82.5$ \\

\noalign{\smallskip}
\hline \noalign{\smallskip}
\end{tabular}
\begin{minipage}[c]{0.97\textwidth}
\footnotesize
\textbf{Notes.} (1): Object name. (2) and (3): The position of the
source in the NIR sky survey, in J2000 coordinates. (4): Spectroscopic
redshift. (5) and (6): \xmm observation ID and date. (7): Total
on-axis exposure times for PN and MOS respectively, before
filtering. (8): Net exposure times at the source position,
accounting for flaring-subtraction and vignetting.  (9):
Source counts at $2$--$10$\,keV, within the source extraction
region ($S$). For J0144 and J1216, the MOS
values are for M2 only. (10): The ratio of net background-subtracted
source counts ($S_{\rm net}$) to the total source counts. Upper limits indicate
non-detections. 
\end{minipage}
\label{xrayData_table}
\end{table*}

\subsection{X-ray data processing and analyses}

\begin{figure*}
	\includegraphics[width=\textwidth]{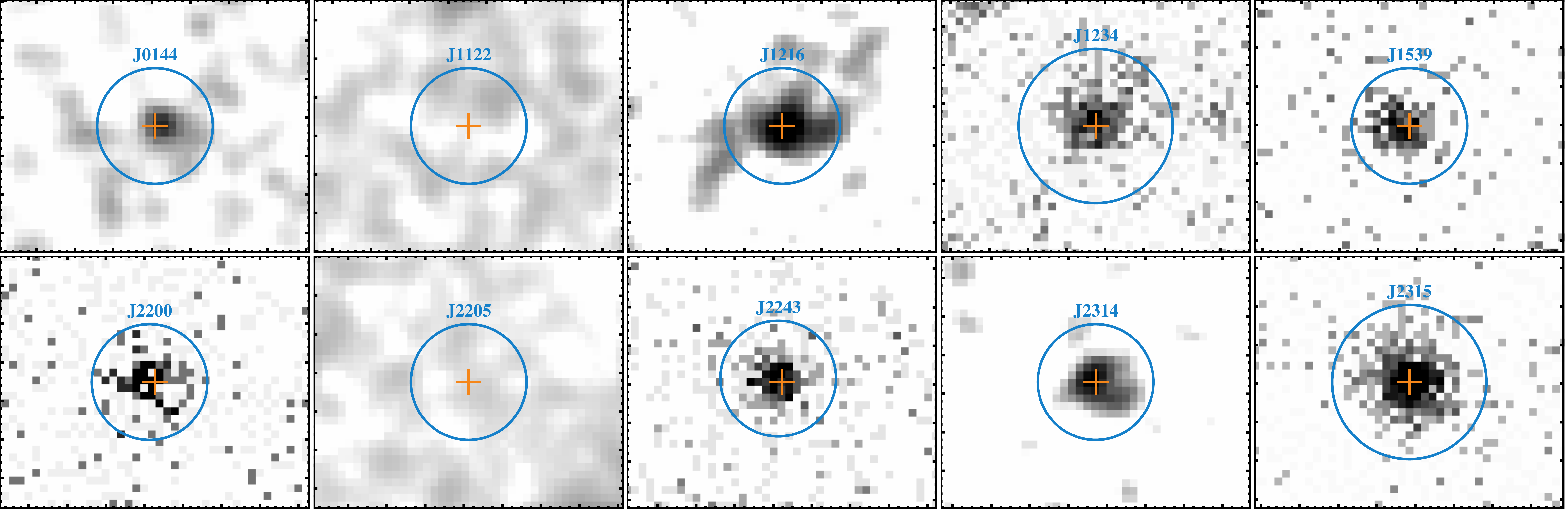}
    \caption{\xmm images at $E=0.5$--$10$~keV from the PN camera for each of the HRQs
      in this work. The near-infrared (NIR) source positions are marked by orange
      crosses, and the extraction regions used for X-ray spectroscopy and
      photometry are shown as blue circles. The images are binned to
      $2$~arcsec per pixel.
      For the five strongest and weakest detections we show unsmoothed
      and smoothed data (with a gaussian of $3$-pixel width), respectively.
      The axes major ticks show
      steps of $10$~arcsec. 
      }
    \label{fig:ds9_fig}
\end{figure*}

We process data from the \xmm Pipeline Processing System, using the
\xmm Science Analysis Software (SAS v17.0.0).
To further clean the data, we examine the background count-rate
lightcurves above $10$~keV and exclude time intervals of significant
background flaring. For PN we make this flaring cut using
rate thresholds between $<0.4$ and $<1$~s$^{-1}$, and for MOS we use
$<0.4$~s$^{-1}$. Table \ref{xrayData_table} lists the net exposure
times following the flaring correction. 

For the photometric and spectral analyses we adopt as standard source regions with
radii of $15$~arcsec, centered on the NIR source position. For the two most strongly detected sources we
use $20$~arcsec. In each case, we extract the background from a
large-area (between $2$ and $18$~sq.~arcmin) source-free region 
nearby, but not contaminated by, the quasar.

\subsubsection{Source detections and photometry}

We produce images in multiple observed-frame and rest-frame energy bands.
For individual images, we extract the total counts from the source
region ($S$) and from the background region ($B$), and apply established aperture photometry
procedures (e.g., \citealt{Lansbury14}) to determine: the binomial
no-source probability ($P_{\rm B}$); the net background-subtracted
source counts ($S_{\rm net}$) and associated Poisson errors; and $3\sigma$ upper limits to $S_{\rm
  net}$ for non-detections (\citealt{Kraft91}).  

  Seven out of ten sources are clearly
  strong detections in the \xmm images (Figure
  \ref{fig:ds9_fig}). Table \ref{xrayData_table} lists the measured
  values of $S$ and $S_{\rm net}/S$. For subsequent analyses of
  these sources, we limit to the energy
  ranges where there is significant signal
  above the background: $0.2$--$10$~keV for J1234, J2200, and J2315;
  $0.2$--$4.5$~keV for J2243; $0.5$--$10$~keV for J1216 and J1539; and
  $0.5$--$4.5$~keV for J2314.
  We constrain the X-ray properties (e.g., luminosities) of these
  sources using spectral modelling (Sections \ref{spectro} and \ref{xray_props}).

Of the remaining three out of ten sources, two (J1122 and J2205) are
undetected at all energies, and one (J0144) has a weak detection. 
J0144 is confirmed to be significantly detected in the $\approx
  0.5$--$3$~keV sub-band ($P_{\rm B}\ll 1\%$), but not at higher
  energies. We characterise these
  three sources using photometry, as spectral modelling is not feasible.
To convert the $S_{\rm net}$ constraints to count
rates, we divide by 
exposure times drawn from exposure maps produced using {\sc eexpmap}, thus accounting for spatially
variable sensitivity (e.g., vignetting). We divide by encircled
energy fractions determined from
{\sc psfgen} simulations, to correct to the $100\%$ PSF flux and to
the account for off-axis effects. To convert from count rates to
fluxes, we use {\sc pimms}-based conversion factors which assume a
power-law spectrum with a photon index of $\Gamma=1.7$ and a
Galactic column density of $N_{\rm H}=3\times 10^{20}$~\nhunit. To
obtain X-ray luminosity constraints in the rest-frame $2$--$10$~keV band, we
perform the above photometry in the appropriate observed-frame band
(e.g., $0.57$--$2.9$~keV for J0144, at $z=2.5$), thus eliminating the
need for extrapolations between bands. 
We also performed photometry at higher rest-frame energies
(e.g., $10$--$40$~keV) but found that the additional information
(e.g., upper limits and band ratios) obtained do not significantly
improve the characterisation of these 
three sources.

\subsubsection{Spectroscopy}
\label{spectro}

For the spectral modelling of the seven well detected sources, we
extract spectral products (spectra and response files) using the source and background
regions above. The individual MOS spectra are combined using {\sc epicspeccombine}.
We group the spectra by a minimum of one count per bin, and model them
using {\sc xspec} (version 12.10.1h; \citealt{Arnaud96}), using the $W$-statistic for fitting. This is
recommended for the low counts regime, and gives results consistent
with $\chi^{2}$ fitting at high counts. We model the PN and MOS
spectra simultaneously.
  We fit the spectra with simple
  absorbed power law models that account for photoelectric absorption
  and Compton-scattering out of the line-of-sight ($\mathtt{cabs\cdot
    zwabs\cdot pow}$ in
  {\sc xspec} formalism).
  We account for the low-level absorption
  through the Galaxy with additional multiplicative absorption
  components fixed to values from \citet{Kalberla05}.
  
 Since the column densities (Section \ref{results}) appear to be low,
  the X-ray spectra are dominated by the transmitted continuum, and
  therefore additional light from Compton scattering by distant material is not expected
  to contribute significantly.
  Therefore, fitting with 
  more detailed physically motivated (e.g., torus) models is unlikely to
  change the main parameter constraints. Nevertheless,
  we perform additional fitting using the torus model of
  \citet{Balokovic18} in a standard setup with the toroidal column
  density uncoupled from the line-of-sight one.

\section{Results and Discussion}
\label{results}

\subsection{X-ray properties}
\label{xray_props}

Given the high redshifts, and the \xmm sensitivity window, it has been
  possible to access energies of $E\approx 0.7$--$35$\,keV in the source
  rest-frame. This means that the relatively unabsorbed high-energy part of the
X-ray spectrum lies within the observed band, and the intrinsic
accretion power and spectral shapes of the quasars can be reliably measured.

\begin{figure*}
	\includegraphics[width=\textwidth]{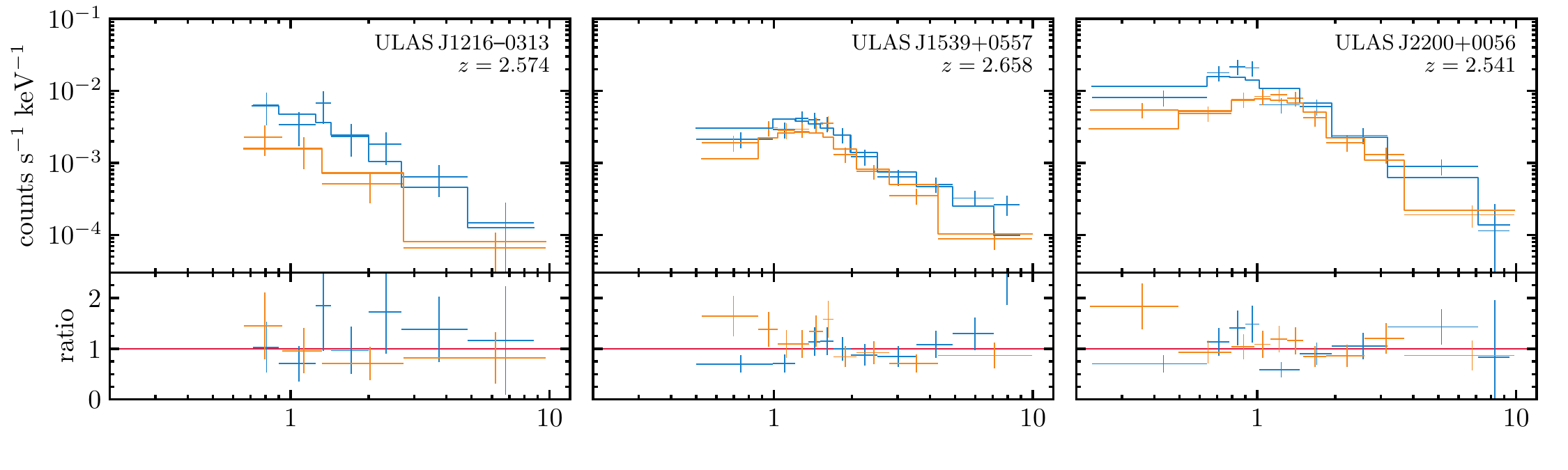}
	\includegraphics[width=\textwidth]{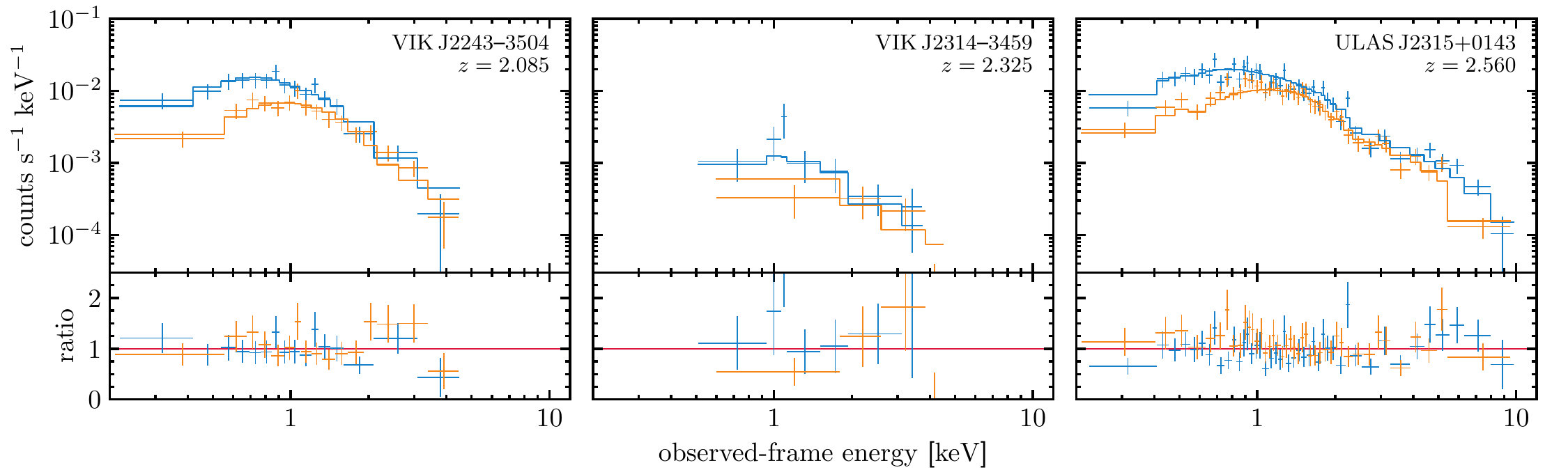}
    \caption{X-ray spectra and data/model ratios for six HRQs with sufficient counts for
      spectral modelling. The seventh such object, J1234, was
      presented in \citet{Banerji14}. The photon energies shown are in
      the observed frame. Given the redshifts of the objects, the full
      range of rest-frame energies sampled is $0.7$--$35$~keV.  The best-fitting absorbed
      power-law models are shown as solid lines, binned to match the
      data. For visual purposes, the data are binned by $2\sigma$ per
      bin for J1216 and J2314, and $4\sigma$ for the remainder. Data
      from the PN and MOS cameras are coloured as blue and orange, respectively.
      }
    \label{fig:lda}
\end{figure*}

  For the seven strongly detected sources, we find that the spectra
  are well described by simple absorbed power law models with no
  additional components (see Table
  \ref{xrayFits_table} and Figure \ref{fig:lda}). In six cases, it is possible to
  simultaneously constrain \nh and the intrinsic photon index
  ($\Gamma$). In the seventh case (J2314) $\Gamma$ was fixed to 
  $1.9$, consistent with the average for the sample. 
  The measured column densities are in the range $N_{\rm
    H}=(1.0$--$4.2)\times 10^{22}$~\nhunit (Table
  \ref{xrayFits_table}).
  For J1234, J1539, and J2315 the column densities measured are in 
  agreement with previous works, which also used simple
    absorbed power law models, within the uncertainties
  (\citealt{Banerji14,Martocchia17}). 
  For one source (J1216), the fitting only places an upper limit on
  \nh. This source and J0144 are both
  detected off-axis, and have small net exposure times; their
  physical constraints would be significantly improved with further \xmm
  observations of modest exposure time. 

  The intrinsic (i.e., unabsorbed) power-law spectra of the modelled HRQs have
  photon indices which appear broadly consistent with the typical range for
  AGNs (i.e., $\bar{\Gamma}\approx 1.8$, $\sigma_{\Gamma}\approx 0.2$;
  e.g., \citealt{Alexander13,Rivers13,Ricci17d}). The mean intrinsic photon
  index for the HRQs with constrained values is $\bar{\Gamma}=1.88$. The most significant
  outlier is J2243, with $\Gamma= 2.5\pm 0.3$. The intrinsic
  luminosities in the rest-frame $2$--$10$~keV band cover the range $\log
  (L_{\rm 2\mbox{-}10\,keV}/\mathrm{erg\ s^{-1}})=44.4$--$45.6$, with
  a median of $45.1$. This confirms that the central engines of HRQs
  are extremely powerful. 
  As shown in Section \ref{lum_section}, when comparing to
    some of the most luminous quasars known in the Universe, the X-ray
    luminosities of the HRQs are in most cases comparably high.

  Fitting instead with the physically motivated toroidal model of
  \citet{Balokovic18}, the best-fit results for \nh, 
  $\Gamma$, and \lx agree closely with those
  above, within the errors, and there is
  no statistical improvement in the fits. Since the best-fit models are 
  dominated by the primary transmitted X-ray continuum, there are no
  informative constraints on toroidal or geometric parameters.

\begin{table}
\centering
\caption{X-ray properties}
\begin{tabular}{llllll}
\hline\hline 
Object & $\Gamma$ & \nh & $L_{\rm X}^{\rm obs}$ & $L_{\rm X}^{\rm int}$ & $W/n$  \\
(1) & (2) & (3) & (4)  & (5)  & (6) \\
\hline 
\noalign{\smallskip}
J0144 & $\cdots$ & $1.4^{\dagger}$ & $44.7$ & $44.7^{\dagger}$ & $\cdots$ \\
J1122 & $\cdots$ & $2.3^{\dagger}$ & $<44.3$ & $<44.3^{\dagger}$ & $\cdots$ \\
J1216 & $1.7^{+0.3}_{-0.3}$ & $<8.0$ & $45.3$ & $45.3$ & $50/54$ \\
J1234 & $1.8^{+0.1}_{-0.1}$ & $2.1^{+0.8}_{-0.7}$ & $45.0$ & $45.1$ & $499/572$ \\
J1539 & $1.6^{+0.2}_{-0.1}$ & $4.2^{+1.7}_{-1.5}$ & $45.0$ & $45.1$ & $332/386$ \\
J2200 & $1.9^{+0.2}_{-0.1}$ & $1.0^{+0.5}_{-0.5}$ & $45.4$ & $45.5$ & $294/335$ \\
J2205 & $\cdots$ & $1.2^{\dagger}$ & $<44.0$ & $<44.1^{\dagger}$ & $\cdots$ \\
J2243 & $2.5^{+0.2}_{-0.2}$ & $1.8^{+0.4}_{-0.3}$ & $45.1$ & $45.1$ & $336/469$ \\
J2314 & $[1.9]$ & $3.8^{+3.0}_{-2.1}$ & $44.3$ & $44.4$ & $99/120$ \\
J2315 & $1.8^{+0.1}_{-0.1}$ & $1.3^{+0.2}_{-0.2}$ & $45.5$ & $45.6$ & $803/888$ \\

\noalign{\smallskip}
\hline \noalign{\smallskip}
\end{tabular}
\begin{minipage}[c]{0.43\textwidth}
\footnotesize
\textbf{Notes.} (1) Short object name. (2): Photon index for the
intrinsic (i.e., unabsorbed) X-ray power law spectrum, and the associated
$1\sigma$ errors. (3): Gas column
density in units of $10^{22}$~\nhunit, and $1\sigma$ errors. (4) and (5):
Logarithm of the observed and intrinsic luminosities in the rest-frame
$2$--$10$~keV energy band. (6) Ratio of the $W$-statistic value to the
number of degrees of freedom. $^{\dagger}$: Estimates based on the average \ebv/\nh.
\end{minipage}
\label{xrayFits_table}
\end{table}

\subsection{Dust versus gas}
\label{dust_gas}

\begin{figure}
	\includegraphics[width=\columnwidth]{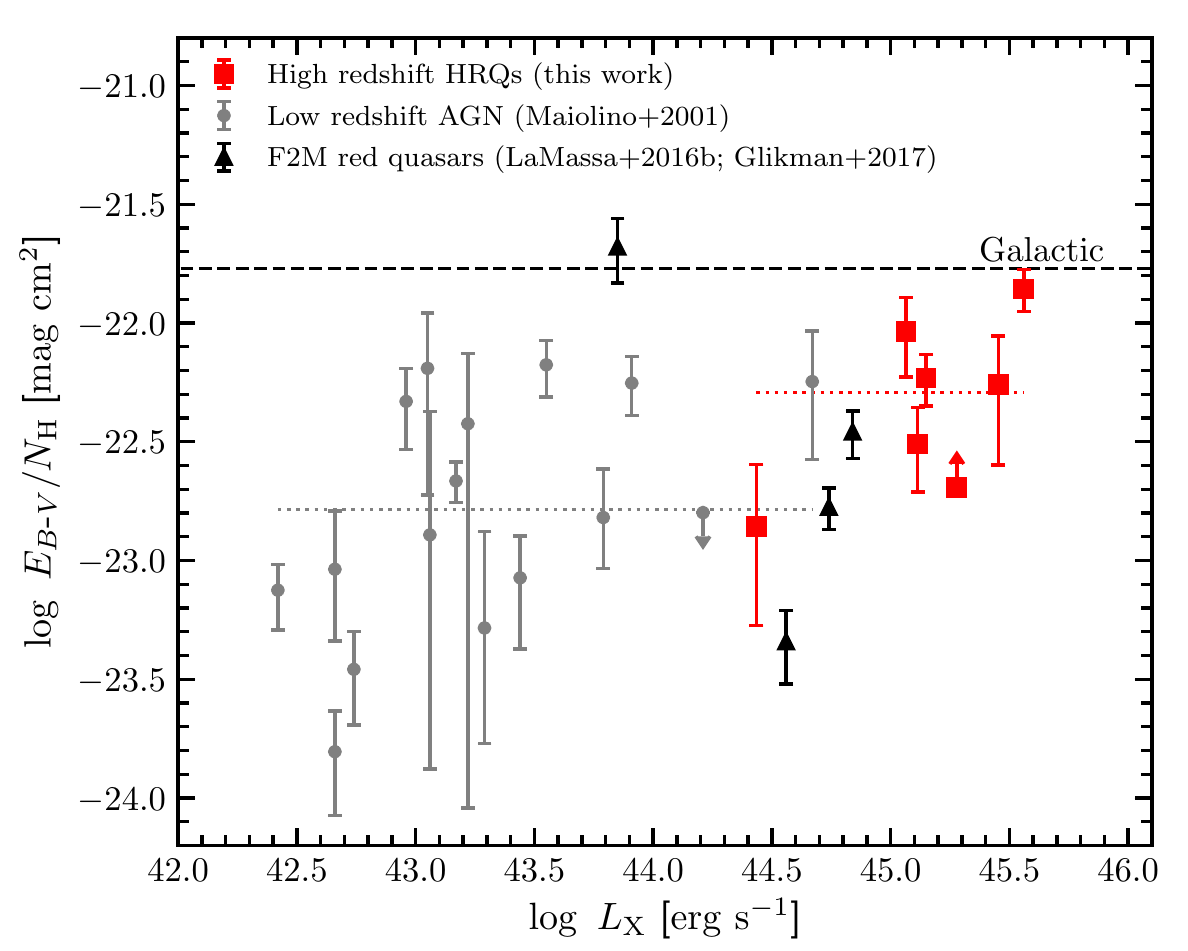}
    \caption{The ratio of dust-reddening ($E_{B-V}$) to gas
      column density (\nh) as a function of intrinsic $2$--$10$~keV 
      luminosity (\lx). The 
      dashed line shows the ratio found for the Galaxy
      (\citealt{Savage79,Guver09}). Dotted lines show the average ratios for
      for lower luminosity AGNs at $z< 0.7$
      (grey; \citealt{Maiolino01}) and for our luminous $z\approx 2.5$ HRQs (red). We also compare with $z\approx 0.5$ F2M
      red quasars (black data points;
      \citealt{LaMassa16c,Glikman17a}). For the HRQs and red quasars, we
    include both the measurement uncertainty on \nh and an assumed
    uncertainty of $\pm 0.15$ for $E_{B-V}$.
      }
    \label{fig:ebvnh_lx}
\end{figure}

Comparison of the gaseous column
density (\nh) with the dust reddening ($E_{B-V}$) can provide
clues as to the nature of the absorbing media in the HRQs.
In Figure \ref{fig:ebvnh_lx} we plot the ratios of $E_{B-V}/N_{\rm H}$ for the seven HRQs with
strong X-ray detections.
The $E_{B-V}$ values were measured
  from the NIR photometry by \citet{Temple19} for eight objects, and
  by \citet{Banerji12} for J1539. For the tenth object, J2315, we adopt the value from
  \citet{Wethers18} based on fitting of the optical-NIR SED.
We find that the ratios range between
$10^{-22.9}$ and $10^{-22.0}$, with an average of $E_{B-V}/N_{\rm H}
=10^{-22.3}$. 
For comparison, we also plot the Galactic value of $E_{B-V}/N_{\rm H}
=10^{-21.8}$ (\citealt{Savage79,Guver09}).  
Although $\approx 3$ sources are consistent with the Galactic value,
there is a systematic offset suggesting that
HRQs have slightly lower dust to gas ratios
than the Galaxy. 

Despite this offset, the HRQs in fact deviate less from the Galactic value than
local AGNs (e.g., \citealt{Maiolino01,Burtscher16,Shimizu18}; see Figure
\ref{fig:ebvnh_lx}) and compared to some other
luminous high-redshift quasars (see Section \ref{other_samples}).
Four other red quasars at $z\approx 0.1$--$0.7$ (selected from FIRST and
  2MASS; ``F2M'' quasars hereafter) studied in the X-ray band by
  \citet{LaMassa16c} and \citet{Glikman17a} have a distribution of
  $E_{B-V}/N_{\rm H}$ values broadly consistent with our HRQ sample
  (black points in Figure \ref{fig:ebvnh_lx}), although the average may be slightly lower.
For the local AGNs (grey points in Figure \ref{fig:ebvnh_lx}), the
highly sub-Galactic $E_{B-V}/N_{\rm H}$ values typically observed are
thought to result, at least partly, from excess X-ray absorption due to
neutral dust-free gas in the innermost $\approx$sub-parsec regions (e.g.,
\citealt{Merloni14,Davies15,Burtscher16,Shimizu18,Liu18}).
One interpretation for the HRQs and the F2M red quasars, therefore, is that they are less
dominated by absorption from the innermost regions. 
The HRQ obscurers then have a higher dust content, with the dust and
gas columns associated with material on similar physical scales,
potentially at larger radial distances from the central black hole
($\gtrsim 100$~pc).
Dust in HRQs has previously been suggested to be
distributed on large scales, based on the narrow-line region (NLR)
properties (\citealt{Temple19}) and the identification of galaxy-scale
dust emitting regions in a few HRQs (\citealt{Banerji18}).
A variability study, using multi-epoch X-ray observations,
would help to better establish the location of the gaseous obscuring media.

\subsection{X-ray versus infrared and bolometric luminosities}
\label{lum_section}

\begin{figure}
\centering
	\includegraphics[width=\columnwidth]{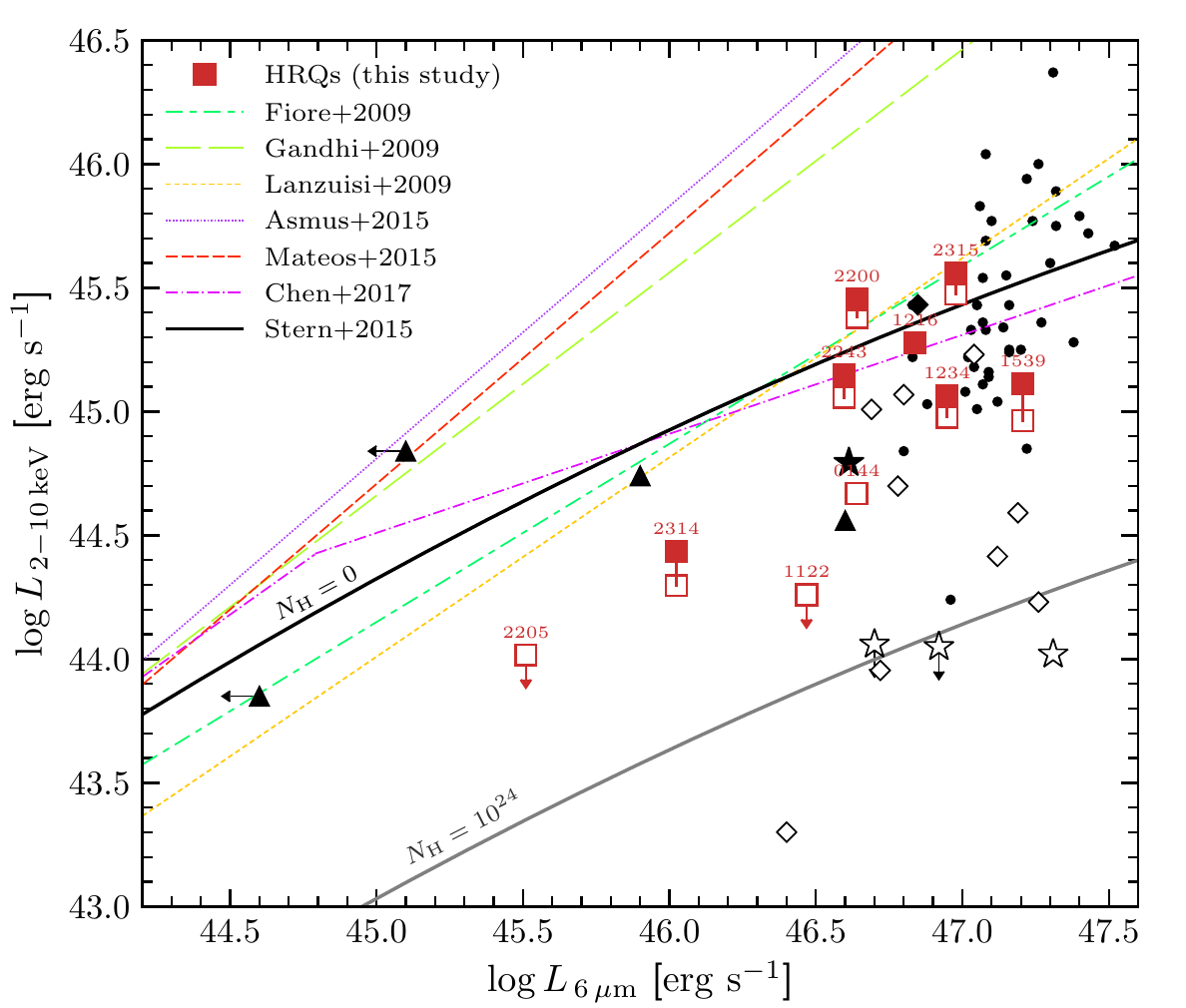}
    \caption{X-ray luminosity versus mid-infrared luminosity, for the
      HRQs (red squares) and other luminous quasar samples. Open
      shapes show absorbed X-ray luminosities and filled shapes show
      intrinsic (i.e., absorption-corrected) X-ray luminosities,
      obtained from X-ray spectral analyses. We compare with three
      other luminous reddened quasar samples: F2M red quasars
      (triangles; \citealt{LaMassa16c,Glikman17a}); extremely red
      quasars (ERQs) from the SDSS
      (diamonds; \citealt{Goulding18}); and {\it WISE} hot~DOGs (stars;
      \citealt{Stern14,Vito18b}). For the latter two samples  
      the intrinsic luminosities shown are averages from the analysis
      of stacked spectra. We also compare with a sample of the most luminous
      type~1 quasars known (small black circles; \citealt{Martocchia17}), and 
      intrinsic \lx--\Lsixum relations from a number of different AGN 
      studies (as labelled). For the luminosity dependent
      relation of \citet{Stern15}, we additionally show a modification of the
      relation (thick gray line) which shows the downscaling in X-ray
      luminosity expected for obscuration by Compton-thick gas 
      ($N_{\rm H}=10^{24}$~\nhunit) along the line of sight.}
    \label{fig:lx_lmir}
\end{figure}

The NIR-selected HRQs have extremely high infrared
and bolometric luminosities (medians of $\left<L_{\rm 6\mu m}\right>=
10^{46.6}$ and $\left<L_{\rm bol}\right>=
10^{47.5}$~\ergpersec, respectively). 
Here we compare these to the measured X-ray
luminosities ($\left<L_{\rm 2\mbox{-}10}\right>=
10^{45.1}$~\ergpersec) in order to test scaling relations at the
highest luminosities in the AGN population, and to better understand
the HRQs in the context of other powerful quasar samples.

The X-ray emission from AGNs arises from a corona, close to
the innermost region of accretion, whilst mid-infrared emission is
thought to arise from distant circumnuclear dust which reprocesses accretion
disk photons. The X-ray
luminosities and the mid-infrared (e.g.,
\sixum) luminosities of AGNs are known
to follow a correlation, which has been characterised using
many different samples (e.g., Figure
\ref{fig:lx_lmir}; 
\citealt{Lutz04,Fiore09,Gandhi09,Lanzuisi09,Levenson09,Ichikawa12,Weedman12,Asmus15,Mateos15,Stern15,Carrera17,Chen17,Toba19}). 
Importantly, when the highest
luminosities are considered ($L_{\rm 6\mu m} \gtrsim
45.5$~\ergpersec), there appears to be deviation from a constant ratio
of \lx:\Lsixum, such that the most luminous objects show relatively
weak X-ray emission.  
 This has lead to the calibration of
luminosity-dependent relations (e.g.,
\citealt{Stern15,Chen17}).

Figure \ref{fig:lx_lmir} shows our HRQ sample compared to various
measurements of the \lx--\Lsixum relation and to other luminous quasar
samples, both reddened and unobscured. For the seven sources with constraints from
X-ray spectral modelling, we are able to show the 
intrinsic (i.e., absorption corrected) X-ray luminosities. For the
remaining three sources (two of which are non-detections), we only
show constraints on the observed (i.e., absorbed) X-ray luminosities.
The intrinsic X-ray luminosity constraints for these three sources should
however lie close to the observed X-ray luminosities, assuming
their dust to gas ratios are similar to the overall sample, and their \nh
values thus lie close to the estimates in Table \ref{xrayFits_table}. Alternatively, these
sources (J1122 and J2205 in particular) could be outlying highly obscured
($N_{\rm H}\gtrsim 10^{23}$~\nhunit) HRQs with anomalous low dust to gas
ratios. The rest-frame \sixum luminosities (\Lsixum) shown were determined from
  the {\it WISE} photometry: for eight sources with detections in all {\it WISE}
  photometric bands, \Lsixum was obtained from interpolation between the $W3$ and $W4$
  bands; for the two sources undetected in the $W4$ band (J2205 and
  J2314), \Lsixum was estimated from the $W3$ flux.

Considering the results for the overall sample, the HRQs do indeed
have intrinsic X-ray to infrared luminosity ratios which are significantly lower
than observed for lower redshift, lower luminosity AGN (e.g.,
\citealt{Lutz04,Gandhi09,Asmus15}), but are in better agreement with
ratios that consider high luminosity AGNs (e.g.,
\citealt{Fiore09,Lanzuisi09,Stern15,Chen17,Martocchia17,Goulding18};
but see also \citealt{Mateos15}). 
The luminosity dependence may be connected to, or driven by, the
underlying relationship between X-ray and bolometric optical/UV emission, where
X-ray emission again becomes relatively weak at the highest
luminosities and accretion rates, due to a
saturation or disruption of the X-ray emitting plasma (e.g., \citealt{Ricci17a,Banados18b}).

The HRQs additionally show some evidence for intrinsic X-ray
weakness (e.g., \citealt{Luo14,Teng14,Vito18c}), on average, with respect to
other typical AGNs of the same infrared luminosity. 
 For instance, eight (out of $10$) of the HRQs lie below
the relation of \citet{Stern15}. 
At a given infrared luminosity, their X-ray luminosities 
cover a range of up to $\approx 1$~dex, suggesting diversity in the
X-ray properties of a population selected on
relatively uniform infrared properties.  
We additionally note that the amount of scatter seen 
around the scaling relations is similar to that
seen for hyperluminous unobscured quasars
(e.g., \citealt{Martocchia17}), especially when the three
weakly and un- detected X-ray sources are
excluded (J0144, J1122, and J2205).

\begin{figure}
	\includegraphics[width=\columnwidth]{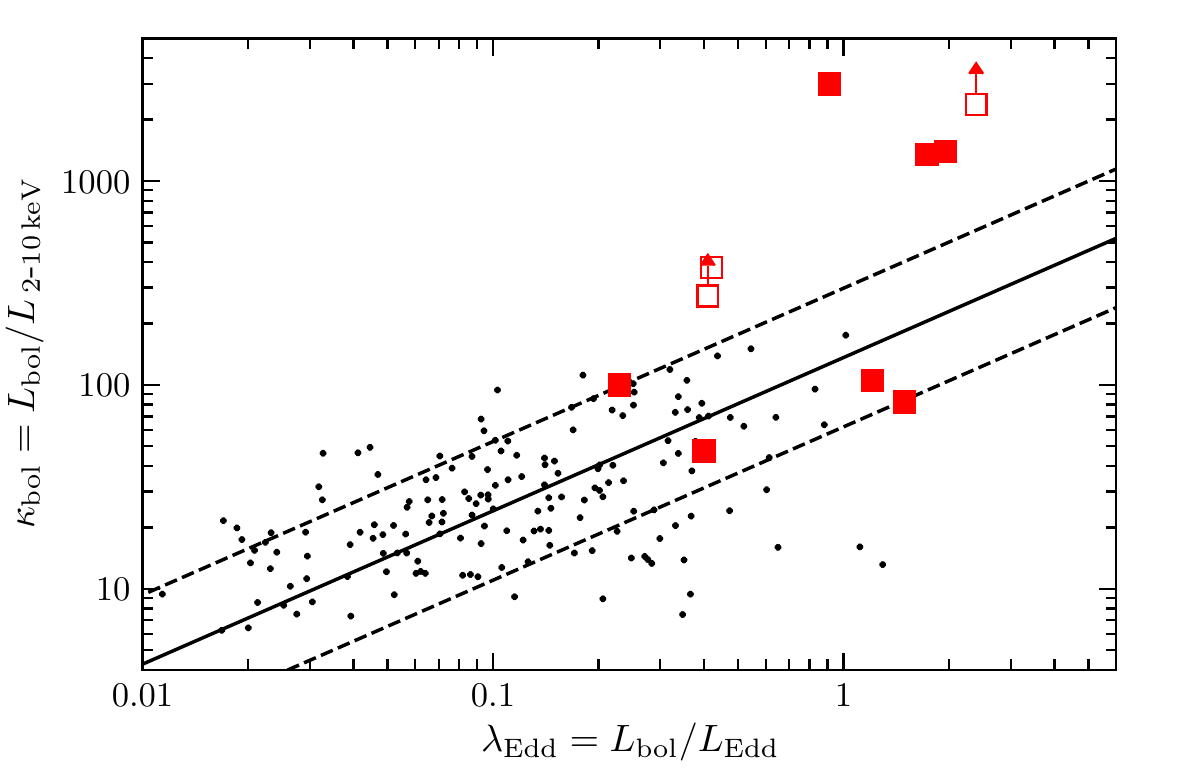}
    \caption{X-ray to bolometric correction ($\kappa_{\rm bol}$) versus Eddington ratio ($\lambda_{\rm
        Edd}$) for the HRQs (red squares). The black points, solid
      line, and dashed lines show Type 1 AGNs, their best fitting
      relation, and the $1\sigma$ dispersion, respectively, from \citet{Lusso12}.
}
    \label{fig:kbol_lbol}
\end{figure}

As mentioned, the X-ray emission of AGNs is also observed to weaken
with increasing accretion power, relative to the bolometric luminosity
inferred from the optical continuum ($L_{\rm bol}$).
This manifests as a positive relation between the X-ray to bolometric
correction ($\kappa_{\rm bol}=L_{\rm bol}/L_{\rm 2\mbox{-}10}$) and the
Eddington ratio ($\lambda_{\rm Edd}=L_{\rm bol}/L_{\rm Edd}$; e.g., \citealt{Elvis94, Vasudevan07, Vasudevan09b, Lusso12}). 
For the HRQs, bolometric luminosities have been estimated from
the optical continuum luminosities at rest-frame $5100$\,\AA\ 
(\citealt{Temple19}). Figure \ref{fig:kbol_lbol} shows our sample in
$\kappa_{\rm bol}$ versus \lamedd, compared to the relation measured for AGNs by \citet{Lusso12}. We find that for four of the HRQs the bolometric
corrections are within the expected range given their Eddington ratios,
with $\kappa_{\rm bol}=47$--$105$. However, the remaining six HRQs
have apparently high values: $\kappa_{\rm bol}\gtrsim 300$ for the
three weakly and un- detected X-ray sources; and $\kappa_{\rm
  bol}=1350$--$2990$ for the strong detections J1216, J1234, and
J1539. The $\kappa_{\rm bol}$ values for J1234 and J1539 are
in agreement with previous estimates from \citet{Banerji14} and
\citet{Martocchia17}, respectively. Overall, the HRQs have
higher bolometric corrections on average than expected from extrapolation of the
$\kappa_{\rm bol}$--$\lambda_{\rm Edd}$ relation for lower luminosity
AGNs.

\subsection{$\lambda_{\rm Edd}$ versus \nh: 
  radiation pressure feedback}
\label{nhlam_section}

\begin{figure}
	\includegraphics[width=\columnwidth]{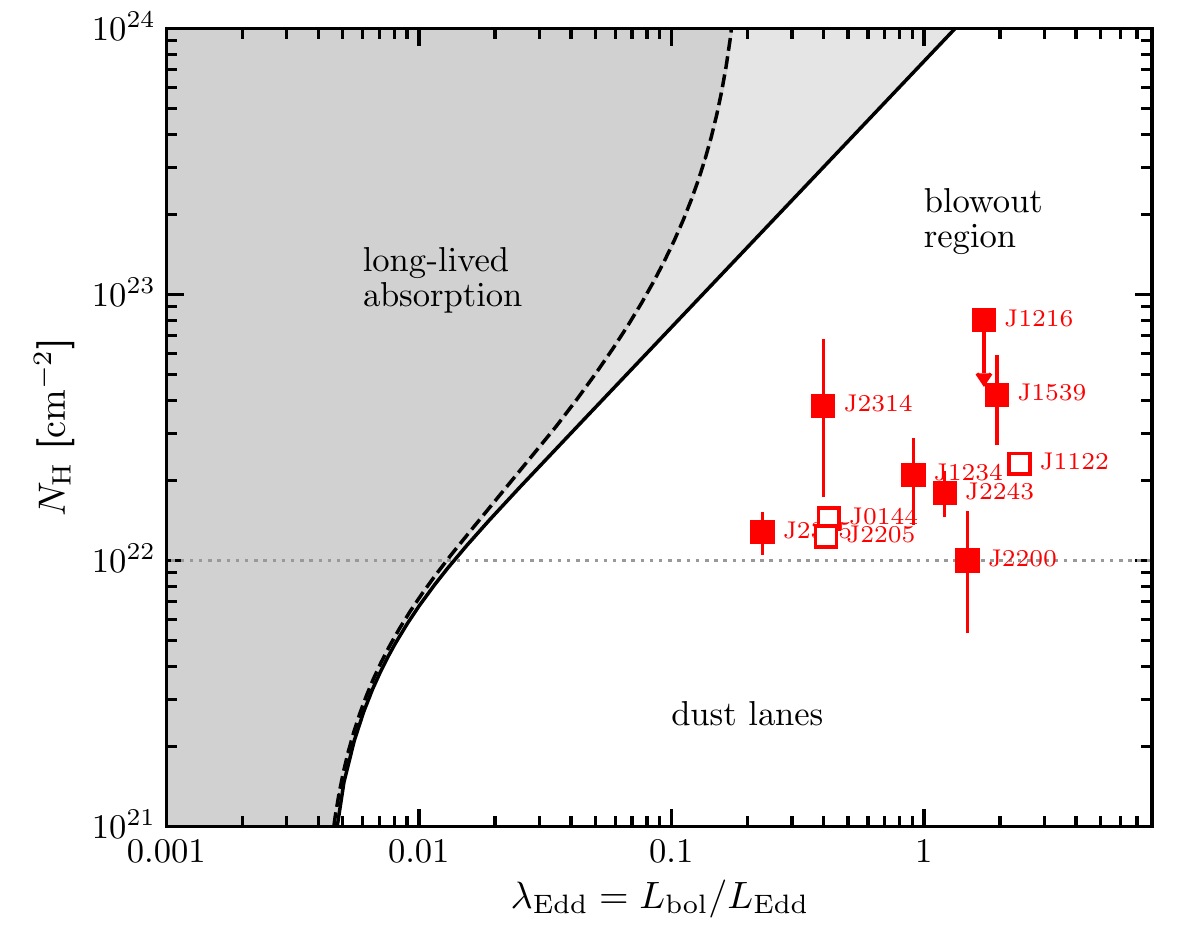}
    \caption{The \nh--$\lambda_{\rm Edd}$ plane for HRQs, i.e.,
      obscuring column density versus Eddington ratio ($\lambda_{\rm
        Edd}=L_{\rm bol}/L_{\rm Edd}$). The tracks show the effective
  Eddington limit for dusty clouds in the cases of single scattering
  (solid line; e.g., \citealt{Fabian09}) and radiation trapping
  (dashed line; \citealt{Ishibashi18b}), above which AGN are predicted
  to undergo a fast blowout phase. Below these limits, obscuring
  clouds may be long-lived.
  The horizontal dotted line marks $N_{\rm H}=10^{22}$~\nhunit, an approximate upper limit to the
column density range expected for 
obscuration through host galaxy dust lanes.  
}
    \label{fig:nh_lambda}
\end{figure}

Our sample of HRQs offers a unique opportunity to examine the
picture of dusty radiative feedback. Firstly, the HRQs have the
extreme luminosities (Section \ref{lum_section}) and dust and gas columns
in their circumnuclear environments (Section \ref{dust_gas})
that are expected for objects undergoing a blowout phase. 
Secondly, the absorption is significant but sufficiently low that the HRQs still have observable
broad lines, which has allowed the measurement of black hole masses
and Eddington ratios from the observed-frame near-IR spectra (\citealt{Temple19}).

Powerful outflows are expected to be launched by black holes accreting
at or above their Eddington luminosity, i.e. the threshold at which 
outward radiative forces match the inward gravitational
pull.
Dust enhances the effect of radiation
pressure around AGNs, due to the effective absorption of UV
light, and the coupling of dust and gas (e.g.,
\citealt{Murray05,Fabian08,Thompson15,Costa18b}). This means that strong outflows are possible even
at luminosities below the classic Eddington limit ($L_{\rm Edd}=1.3\times
10^{38}$\,$M_{\rm BH}/M_{\odot}$), which considers only a simple gas
(e.g., \citealt{Fabian06b,Fabian08,Roth12}). 
In other words, there is an effective
Eddington limit for dusty gas, which is a function of the
circumnuclear gas density (see the
tracks in Figure \ref{fig:nh_lambda}). 

In the low redshift Universe, there is evidence of this mechanism of
radiation pressure feedback at work in AGNs.
Hard X-ray selected AGN 
 appear to preferentially avoid the ``forbidden'' (or ``blowout'')
 region in the \nh--\lamedd plane, above the effective Eddington
 limit, where blowing out of the interstellar medium is expected
 (\citealt{Fabian09,Raimundo10, Vasudevan13,Ricci17c,Bar19}; see also \citealt{Ballo14}). 
\citet{Ricci17c} demonstrated this for a large sample of $392$
\swiftbat AGN ($\left<z\right>=0.037$), and showed that this
physical connection between the obscuring environment of AGN and their
accretion rates can
help to explain observables such as the anticorrelation between obscured
AGN fraction and luminosity (e.g., \citealt{Burlon11}).

In Figure \ref{fig:nh_lambda}, we show the location of the X-ray
observed HRQs on the \nh--\lamedd plane. We compare with tracks showing
the effective Eddington limit for dusty gas, both in the
single scattering limit (e.g., \citealt{Fabian09}) and the
radiation trapping limit (\citealt{Ishibashi18b}). 
Overall, the HRQ sample clearly lies above these limits, and thus in the
blowout region. 
As usual for distant AGN, knowledge of the \lamedd values is limited by the
black hole mass measurements. For these sources the individual virial black hole
masses have uncertainties of $\approx \pm 0.4$~dex (\citealt{Temple19}).
Some individual sources may therefore, in reality, lie closer to the
effective Eddington limits. However, the large overall average offset observed
of $1.5$~dex from the solid line is difficult to explain by uncertainties alone. 
Alternatively, it is possible
that systems appearing above the limit have obscuring material at large
distances, and are thus under the influence of a larger enclosed
gravitational (stellar) mass, in which case the tracks
in Figure \ref{fig:nh_lambda} would move rightwards.
The enclosed mass must be large ($\gtrsim 10
M_{\rm BH}$) for this to solely account for the HRQ data points.

\subsubsection{Evidence for ongoing outflows}

As shown above, the HRQs lie in the realm of \nh-\lamedd parameter space expected to
be conducive to the driving of large-scale outflows via radiative
pressure on dust. It is thus interesting to consider evidence for ongoing outflows. The X-ray spectra (Section
\ref{xray_props}) are not currently sufficient to place constraints on
blueshifted emission around the Fe line complex. Future sensitive X-ray observations with high
spectral resolution (e.g., with {\it Athena})
may be required to pick out such features. However, the 
observed-frame near-infrared spectra of the parent HRQ sample do show evidence for
outflows. Specifically, broad components are
identified in the emission line profiles of \oiii $\lambda 5007$.

Eight of the HRQs in this work have spectroscopic coverage of the \oiii and Balmer
line complexes, which were modelled by \citet{Temple19}. In five cases
\oiii is strongly detected, and in each case significant broadening of
the \oiii line is measured. For four HRQs the $80\%$ velocity
width is large, with a range of $w_{80}=1710$--$2572~\mathrm{km\
  s^{-1}}$, and in two cases (J2200 and J2243) strong blueshifted wings
are clearly observed in the \oiii line profiles. These are
also the two cases with the highest signal-to-noise. 
For the remaining three sources (J0144, J1216, and J1234) \oiii is
undetected, and appears to be intrinsically underluminous in the
latter two sources. This is consistent with a picture where overionization or 
the expulsion of line-emitting gas results in weak emission from the
narrow-line region (e.g., \citealt{Temple19,
  Netzer04}), and also with the finding that intrinsically weak
  \oiii emission is relatively common in high-luminosity high-redshift
  quasars compared to the lower redshift population (e.g., \citealt{Coatman19}).
There is further evidence for outflows in HRQs that currently lack X-ray
coverage. For instance, assuming the average dust-to-gas ratios in Section
\ref{dust_gas}, the two HRQs with the next highest \nh and \lamedd
values (J1117 and J1303) also exhibit very broad \oiii profiles
($w_{80}=2277$ and $2898~\mathrm{km\ s^{-1}}$, respectively). 

In summary, there is evidence for strong outflows of ionised gas around
these HRQs, which may ultimately be connected to their location in
the \nh--\lamedd diagram, thus linking the line kinematics to 
radiative feedback.

\begin{figure*}
      \centering
	\includegraphics[width=0.7\textwidth]{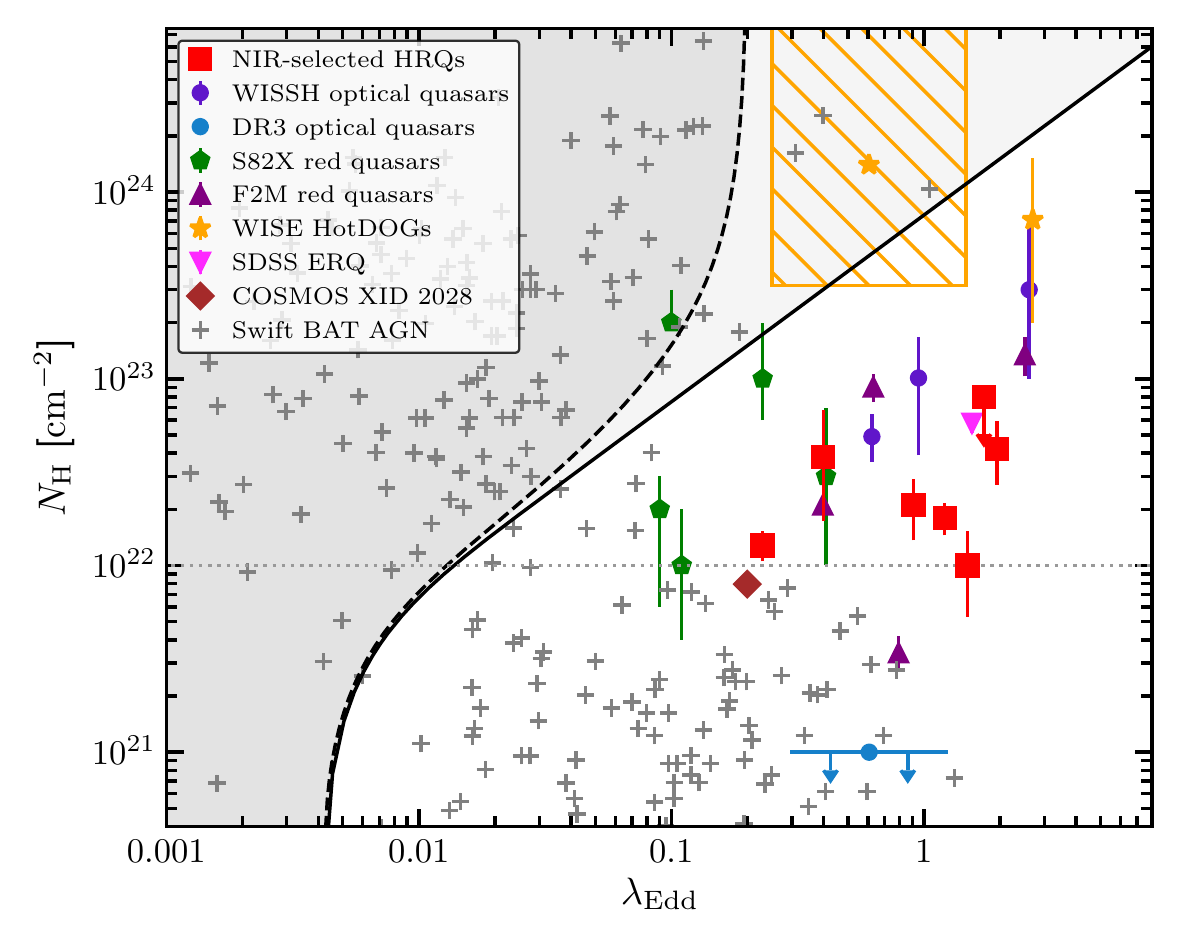}
    \caption{\lamedd versus \nh, as in Figure
      \ref{fig:nh_lambda}. Here we
      compare with multiple different samples, and only show 
      \nh constraints obtained from X-ray spectral modelling. In
      addition to our $z\approx 2.5$ HRQs, multiple other high luminosity
      reddened quasar samples are shown, including:  {\it WISE}
      hot~DOGs at $z\gtrsim 2$
      (hatched orange region, based on \citealt{Vito18b}
      and \citealt{Wu18}); a $z=1.0$ hot~DOG
      (orange star; \citealt{Ricci17a}); F2M red quasars at $z\approx
      0.5$ (purple triangles; \citealt{LaMassa16c,Glikman17a}); $z\approx 1$ red quasars
      in the Stripe 82 X-Ray Survey (green pentagons;
      \citealt{LaMassa17}); an ERQ at $z=2.3$ (magenta inverted triangle;
      \citealt{Goulding18,Perrotta19}); and the $z\approx 1.6$ dust-obscured quasar
      XID~2028 (brown diamond; \citealt{Brusa10,Kakkad16}). For hyperluminous optical quasars we
      show: the core ($\pm 1\sigma$) range of \lamedd values for the
      most luminous quasars at $z \gtrsim 2$
      in SDSS DR3 (blue line, plotted at $N_{\rm H}=10^{21}$~\nhunit);
      and the three $z\approx 2.5$ WISSH quasars mentioned in the text (purple circles;
      \citealt{Martocchia17}). The  ``+'' symbols show local
      ($z\approx 0.037$) \swiftbat
      AGN, which appear to avoid the blowout region
      (\citealt{Ricci17c}; \nh uncertainties not plotted). The latter
      sample also contains many unobscured AGN at low \nh values,
      outside the range shown.
  }
    \label{fig:nh_lambda_all}
\end{figure*}

\subsection{Comparison with other luminous quasar samples}
\label{other_samples}

Here we consider the results in the context of other relevant samples
of highly luminous quasars which have been observed in the X-ray band. 
Figure \ref{fig:nh_lambda_all} shows the \nh-\lamedd plane again, with
the comparison samples plotted. We
additionally compare with the lower luminosity, low-redshift AGN
population selected by \swiftbat (\citealt{Ricci17c}).

\subsubsection{Reddened quasars}

Comparable in luminosity ($L_{\rm bol}\gtrsim 10^{46.6}$~\ergpersec) and redshift to our HRQs are the population
of highly reddened hot dust-obscured galaxies (hot~DOGs), selected
from the mid-infrared {\it WISE}
all-sky survey as ``W1W2 dropouts''. 
Hot~DOGs are so named due to hotter dust temperatures ($T\gg 60$~K)
with respect to ULIRGs, SMGs, and normal DOGs (\citealt{Eisenhardt12,Tsai15,Assef15}).
Similarly to HRQs, hot~DOGs have comparable number densities
to $z\gtrsim 2$ SDSS quasars at the highest luminosities ($L_{\rm
  bol}\gtrsim 10^{47}$~\ergpersec; e.g., \citealt{Assef15}).
The most luminous hot~DOGs lie at $z\approx 2$--$4$, are thought to be
accreting close to the Eddington limit (e.g.,
\citealt{Assef15,Tsai15,Tsai18,Wu18}), and are hidden by material which is
 either Compton-thick or highly obscured and Compton-thin (i.e.,
  $N_{\rm H}\gtrsim 10^{23.5}$~\nhunit; e.g.,
  \citealt{Stern14,Piconcelli15,Assef16,Ricci17a,Vito18b,Zappacosta18b,Assef19}).

  In Figure \ref{fig:nh_lambda_all} we show a hatched region where the 
  hot~DOG population may lie, based on the column densities measured
  by \citet{Vito18b} and assuming the Eddington ratio range measured by
  \citeauthor{Wu18} (\citeyear{Wu18}; for a different subsample where
  broad lines are visible). 
  We point out that hot~DOGs may reside in an interesting region of the
  \nh--\lamedd plane, below the Eddington limit for single scattering,
  but above the Eddington limit for radiation trapping. In other
  words, the high accretion rates of hot~DOGs may lead to blowout due
  to radiative pressure feedback, but only if the infrared optical
  depth is sufficiently high (e.g., $\tau_{\rm IR}\gg 1$;
  \citealt{Ishibashi18b}). 
  The $z\approx 2$ hot~DOGs studied by \citet{Wu18} show some evidence
  for outflows in their optical emission lines. 
  The most luminous hot~DOG WISE\,J2246$-$0526 shows evidence for
  dramatic gas kinematics in the host galaxy, consistent with being
  due an emerging outflow driven by dusty radiation pressure feedback
  (e.g., \citealt{DiazSantos16}). 

Also comparable in luminosity and redshift are the extremely red
quasars (ERQs; \citealt{Ross15,Hamann17}) which have been identified in the
SDSS Baryon Oscillation Spectroscopic Survey (BOSS).
  This population at $z\approx 2$--$3$ is selected based on high
  infrared to optical flux ratios and high equivalent width \civ
  line emission. The sample is inconsistent with simple dust
  reddening, leading to a somewhat uncertain extinction ($E_{B-V}\gtrsim 0.3$; see discussion in \citealt{Hamann17}).
  \citet{Goulding18} presented the X-ray properties of ERQs, finding
  that they are mostly highly obscured or close to Compton-thick
  (e.g., $N_{\rm H}\approx 8\times 10^{23}$~\nhunit), and are likely
  obscured by their own equatorial outflows. 
\citet{Zakamska16b} and \citet{Perrotta19} found dramatic fast \oiii outflows,
reaching velocities up to $\approx 7000\ \mathrm{km\ s^{-1}}$.
We note that many ERQs are likely to lie in the blowout region, based
on their expected \nh and \lamedd values. In Figure
\ref{fig:nh_lambda_all} we plot one ERQ (SDSS
J000610.67+121501.2) which has a constrained \nh (\citealt{Goulding18}) and a \lamedd
measurement from \citet{Perrotta19}.

\citet{Glikman12} presented a large sample of red quasars selected
from FIRST and 2MASS with red optical to near-infrared colours. This
population extends to the values of
dust reddening, luminosity, and redshift found for our HRQs.
Four of the lower redshift F2M quasars, with $z\approx 0.1$--$0.7$, $E_{B-V}=0.6$--$1.5$, and
$L_{\rm 6\mu m}=10^{44.6\mbox{-}46.6}$~\ergpersec, have
been studied with X-ray spectral modelling
(\citealt{LaMassa16c,Glikman17a}).
The range of $E_{B-V}/N_{\rm H}$ values covered by the F2M quasars is
similar to the HRQs.
\citet{Glikman17b} demonstrated that these lower redshift F2M quasars
lie in the blowout region (as shown again here in Figure \ref{fig:nh_lambda_all}), and
have merging host galaxies, thus reinforcing the picture that red
quasars are a transitional radiatively driven blowout phase following a galaxy merger.

Red quasars have also been identified in X-ray surveys. As a notable example,
the brightest obscured quasar in the COSMOS survey, XID~2028
($z=1.592$; $L_{\rm bol}\approx 10^{46.3}$~\ergpersec;
\citealt{Brusa10}), also satisfies
the red quasar criteria of \citet{Banerji12}.
Followup of XID~2028 revealed outflows in multiple gas phases
(e.g., \citealt{Brusa15a, Brusa15b, Brusa16}). \citet{Kakkad16}
showed that the quasar lies close to the blowout region, along with
some other lower luminosity AGN exhibiting \oiii outflows, and we plot
a data point for XID~2028 in Figure \ref{fig:nh_lambda_all} for comparison.

In summary, HRQs and other luminous reddened quasar populations appear to
occupy the forbidden blowout region, and simultaneously display other observational
features (e.g., fast outflows) expected for a blowout phase. This
potentially connects the kinematics in the quasar host galaxies to the
underlying physical mechanism of AGN feedback during phases of
extremely fast accretion onto supermassive black holes.
Deeper, more detailed study of the populations in the blowout region
will continue to help to establish the role of radiative pressure feedback in
influencing galaxy properties, including the importance relative 
to other mechanisms of feedback from winds or jets (e.g., \citealt{Veilleux05,Wagner16,Wylezalek18,Jarvis19}).

\subsubsection{Optical quasars}

In addition to the reddened quasar samples discussed above, we
consider the properties of comparably luminous unobscured (unreddened) quasars. 
The most luminous optical quasars in SDSS DR3 (with $M_{i}<-29.28$)
were presented in \citet{Just07}
and \citet{Stern15}, and include objects at $z\approx 2$--$3.5$, with
$L_{\rm 6\mu m}\approx 10^{46.5\mbox{-}47.5}$~\ergpersec. A data point
is shown in Figure \ref{fig:nh_lambda_all} to highlight the core ($\pm
1\sigma$) range of Eddington ratios measured for these optical
quasars. This is similar to the range of Eddington ratios found for
the HRQs, but optical quasars typically have low intrinsic \nh values (as 
expected from their low dust reddening), meaning they generally lie well below 
red quasars in the \nh-\lamedd diagram.  
Nevertheless, luminous optical quasars do show evidence of outflows. In fact,
\citet{Temple19} find no significant difference between the kinematic \oiii
properties of HRQs and optical quasars when matching in luminosity and 
redshift. It may be that large-scale outflows, initially
launched by radiation pressure during a reddened quasar phase, persist
and remain observable at the time when the central engine is exposed as an
optical quasar.

There are some cases, in exception to the above, of
  luminous optical quasars with relatively high \nh values, and which thus
  appear to lie in the blowout region. In Figure
  \ref{fig:nh_lambda_all} we show, as an example, three
such cases from the WISE/SDSS selected hyper-luminous (WISSH) quasar
sample (\citealt{Weedman12, Bischetti17, Martocchia17}).
These three were identified by taking the X-ray studied WISSH sample
of \citeauthor{Martocchia17} (\citeyear{Martocchia17}; 41 sources overall), and
 taking the sources with: constrained \nh values; and \lamedd
 estimates (from \citealt{Shen11}) which place them in the blowout
 region. The remaining WISSH sources mostly have upper limits in
 column density (i.e., consistency with low \nh).
The three sources shown, SDSS~J132827.06$+$581836.8,
J152156.48$+$520238.4, and J154938.71$+$124509.1, may
 be cases where the high \nh arises from dust-free gaseous
 absorption close to the nucleus. J1328 and J1549 both belong to the
 minority of (nine) WISSH sources studied by \citet{Martocchia17}
 which are highlighted as being broad absorption line (BAL)
 quasars. In such systems, fast outflows occur along the line-of-sight,
 likely originating from winds around 
 the inner accretion disk. The third source, J1521, is extremely
 luminous ($L_{\rm bol}\approx 10^{48.2}$~\ergpersec) and belongs to
 the rare category of weak-line quasars (WLQs). In WLQs, a geometrically thick inner accretion disk is hypothesised to cause
 the gaseous X-ray obscuration (\citealt{Wu11,Luo15}). Considering the
 above, care should be taken when interpreting the location of a given source in
 the \nh--\lamedd plane, without consideration of the broader multiwavelength
 properties, such as dust to gas ratios.

\subsubsection{Time evolution}
\label{time_evolution}

  Generally very few AGNs lie in
  the blowout region. This suggests that flickering to accretion rates above
  the effective Eddington limit is uncommon over short timescales,
  and that blowout (to lower \nh values) occurs relatively quickly.
  These HRQs may therefore represent a brief, but important,
  phase in an evolutionary sequence for massive galaxies
 (e.g.,
 \citealt{Hopkins08,Somerville08,Alexander12,Bridge13}) 
 driven by radiative pressure feedback (\citealt{Ishibashi16b}).
  In this picture, some quasars may graduate downwards in
  Figure \ref{fig:nh_lambda_all}: a deeply enshrouded Type~2 or hot~DOG-like
  phase occurs, with accretion exceeding the effective Eddington limit due
  to a plentiful fuel supply (and hence high-\nh); the expulsion of
  material follows, resulting in a red quasar, and then
  finally an unobscured quasar phase. 
  \citet{Wu18} found this picture to be consistent with the black hole
  properties of these samples at $z\sim 2$. Some further support is
  added  by observations of starbursts and merging host galaxies for the various
  populations (e.g., \citealt{Urrutia08,Glikman15,Fan16a,Banerji18,DiazSantos18}).

  For a rough estimate of the timescale for blowout, given our measurements
  for HRQs, we consider the simplified case of an expanding shell of
  material. 
  We assume that the matter is initially close to Compton-thick, with a column
  density of
  $N_{\rm H,0}=10^{24}$~\nhunit, lies within an
  initial radius of $r_{0}=30$~pc, and thus has a mass of $M\approx 4\times 10^{7} M_{\odot}$.
Assuming that the column density evolves as $N_{\rm H}\propto r^{-2}$
(e.g., \citealt{Ishibashi15}),\footnote{The evolution can be flatter,
  for instance, if this is the first outflowing shell and a significant amount of the ISM is swept up.} 
the radius must increase by a factor of seven to
reach the column densities observed for HRQs ($N_{\rm H}\approx
2\times 10^{22}$~\nhunit; Section
\ref{xray_props}). It follows that the timescale for this phase of the
blowout is $t\approx 7\times (3\times 10^{4}\,\mathrm{yr}) \times
(\frac{r_{0}}{30\,\mathrm{pc}})
(\frac{v}{1000\,\mathrm{km\,s^{-1}}})^{-1}$.
In this simple picture, the obscuring material in HRQs
should lie at radii of
order $\sim 200\,\mathrm{pc}$. This is in agreement with our results for
the dust-to-gas ratios (Section \ref{dust_gas}), which suggest that the obscuration
occurs on larger physical scales than in typical AGN.
  Future study should address how well this timescale for blowout agrees with the fraction of
  sources in the forbidden region, and the space densities of HRQs and
  other luminous quasars.

\section{Summary}
\label{summary}

We have reported on \xmm observations of ten heavily dust-reddened
quasars (HRQs), five of which are from recent new dedicated observations. 
This has allowed us to characterise the
high-energy properties of HRQs at $\left< L_{\rm bol} \right>=
10^{47.5}$~\ergpersec and $\left< z \right>= 2.5$. These represent an
important population of quasars, which are similar in number density
to the most
luminous unobscured (e.g., SDSS) quasars at these redshifts.
Due to the high redshifts of the sources, we have been able to access energies of
$E \approx 0.7$--$35$\,keV in the source rest-frame, a
range where the primary X-ray coronal emission is well sampled. 
Our main results are as follows:

\begin{itemize}

\item Of the five newly observed HRQs, three are well detected (J2200,
  J2243, and J2314) and two are
  undetected (J1122 and J2205) by \xmm at $0.5$--$10$~keV. 
  Including the other $z>2$ HRQs with X-ray coverage, this brings the total numbers
  of X-ray detected and undetected HRQs to eight and two, respectively;
  see Section \ref{sample_data}. 

\item   The measured gas column densities are in the range $N_{\rm H}=(1$--$8)\times
  10^{22}$~\nhunit, and the typical intrinsic (i.e., unabsorbed) photon index is found to
  be $\bar{\Gamma}\approx 1.9$.
  The intrinsic X-ray luminosities measured are $L_{\rm
  2\mbox{-}10\,keV} = 10^{44.4}$--$10^{45.6}$~\ergpersec, with an
  average of $10^{45.1}$, confirming these as powerful AGNs in
  the X-ray band; see Section \ref{xray_props}.

\item We explore the dust to gas ratios in the extreme environments of
  the HRQs by determining $E_{B-V}/N_{\rm H}$. 
The ratios measured (average of $E_{B-V}/N_{\rm H} =10^{-22.3}$)
suggest a relatively high dust content for the obscurers, compared
to other AGNs; see Section \ref{dust_gas}.

\item The intrinsic X-ray luminosities ($L_{\rm 2\mbox{-}10\,keV}$)
  measured are weaker, relative to \Lsixum and \Lbol, than is
  found for lower redshift, lower luminosity AGN. Broadly, this agrees
with other hyperluminous quasar samples. However, the HRQs show tentative
evidence for being even more (intrinsically) X-ray weak than other quasars of
comparable luminosity. The X-ray to
bolometric corrections reach up to extremely high values of
$\kappa_{\rm bol}>1000$; see Section \ref{lum_section}.

\item We find that all of the HRQs are in agreement with moderate
  column densities and high Eddington ratios ($\lambda_{\rm
    Edd}>0.2$), placing them in a region of the \nh--$\lambda_{\rm
    Edd}$ plane which is systematically avoided by X-ray survey
  samples (e.g., \citealt{Ricci17c}). In this ``forbidden'' or
  ``blowout'' region, radiative pressure feedback acting on
  dusty gas is expected to drive strong outflows. In agreement with
  this picture, the narrow-line region (\oiii) properties of the HRQs
  studied here show evidence for ongoing outflows; see Section \ref{nhlam_section}.

\item Along with the HRQs, a number of other highly luminous quasar
  samples appear to lie in the blowout region for radiative pressure
  feedback. 
  However, we note that the wider source
  properties (e.g., dust to gas ratios) must be considered when
  interpreting this.
  The locations of different samples in the \nh--$\lambda_{\rm
    Edd}$ plane are broadly compatible with a picture where different 
  populations are connected by an evolutionary
  sequence; see Section \ref{other_samples}.

\end{itemize}

\section*{Acknowledgements}

This work was supported by a Herchel Smith Research
Fellowship of the University of Cambridge (G.B.L.); and ERC Advanced
Grant 340442 (A.C.F.). We thank the anonymous referee for
  their review, which improved this work.
Paul Hewett, Wako Ishibashi, Roberto Maiolino, and Richard
McMahon are thanked for the useful discussions.
This work is based on observations obtained with \xmm, an
ESA science mission with instruments and contributions directly funded
by ESA Member States and NASA.




\bibliographystyle{mnras}
\bibliography{bibliography.bib}{}

\begin{thebibliography}{}
\makeatletter
\relax
\def\mn@urlcharsother{\let\do\@makeother \do\$\do\&\do\#\do\^\do\_\do\%\do\~}
\def\mn@doi{\begingroup\mn@urlcharsother \@ifnextchar [ {\mn@doi@}
  {\mn@doi@[]}}
\def\mn@doi@[#1]#2{\def\@tempa{#1}\ifx\@tempa\@empty \href
  {http://dx.doi.org/#2} {doi:#2}\else \href {http://dx.doi.org/#2} {#1}\fi
  \endgroup}
\def\mn@eprint#1#2{\mn@eprint@#1:#2::\@nil}
\def\mn@eprint@arXiv#1{\href {http://arxiv.org/abs/#1} {{\tt arXiv:#1}}}
\def\mn@eprint@dblp#1{\href {http://dblp.uni-trier.de/rec/bibtex/#1.xml}
  {dblp:#1}}
\def\mn@eprint@#1:#2:#3:#4\@nil{\def\@tempa {#1}\def\@tempb {#2}\def\@tempc
  {#3}\ifx \@tempc \@empty \let \@tempc \@tempb \let \@tempb \@tempa \fi \ifx
  \@tempb \@empty \def\@tempb {arXiv}\fi \@ifundefined
  {mn@eprint@\@tempb}{\@tempb:\@tempc}{\expandafter \expandafter \csname
  mn@eprint@\@tempb\endcsname \expandafter{\@tempc}}}

\bibitem[\protect\citeauthoryear{{Aird}, {Coil}, {Georgakakis}, {Nandra},
  {Barro}  \& {P{\'e}rez-Gonz{\'a}lez}}{{Aird} et~al.}{2015}]{Aird15a}
{Aird} J.,  {Coil} A.~L.,  {Georgakakis} A.,  {Nandra} K.,  {Barro} G.,
  {P{\'e}rez-Gonz{\'a}lez} P.~G.,  2015, \mn@doi [\mnras]
  {10.1093/mnras/stv1062}, \href
  {http://adsabs.harvard.edu/abs/2015MNRAS.451.1892A} {451, 1892}

\bibitem[\protect\citeauthoryear{{Alexander} \& {Hickox}}{{Alexander} \&
  {Hickox}}{2012}]{Alexander12}
{Alexander} D.~M.,  {Hickox} R.~C.,  2012, \mn@doi [\nar]
  {10.1016/j.newar.2011.11.003}, \href
  {http://ukads.nottingham.ac.uk/abs/2012NewAR..56...93A} {56, 93}

\bibitem[\protect\citeauthoryear{{Alexander} et~al.,}{{Alexander}
  et~al.}{2013}]{Alexander13}
{Alexander} D.~M.,  et~al., 2013, \mn@doi [\apj] {10.1088/0004-637X/773/2/125},
  \href {http://adsabs.harvard.edu/abs/2013ApJ...773..125A} {773, 125}

\bibitem[\protect\citeauthoryear{{Arnaud}}{{Arnaud}}{1996}]{Arnaud96}
{Arnaud} K.~A.,  1996, in {Jacoby} G.~H.,  {Barnes} J.,  eds,  Astronomical
  Society of the Pacific Conference Series Vol. 101, Astronomical Data Analysis
  Software and Systems V. p.~17

\bibitem[\protect\citeauthoryear{{Asmus}, {Gandhi}, {H{\"o}nig}, {Smette}  \&
  {Duschl}}{{Asmus} et~al.}{2015}]{Asmus15}
{Asmus} D.,  {Gandhi} P.,  {H{\"o}nig} S.~F.,  {Smette} A.,   {Duschl} W.~J.,
  2015, \mn@doi [\mnras] {10.1093/mnras/stv1950}, \href
  {http://adsabs.harvard.edu/abs/2015MNRAS.454..766A} {454, 766}

\bibitem[\protect\citeauthoryear{{Assef} et~al.,}{{Assef}
  et~al.}{2015}]{Assef15}
{Assef} R.~J.,  et~al., 2015, \mn@doi [\apj] {10.1088/0004-637X/804/1/27},
  \href {http://adsabs.harvard.edu/abs/2015ApJ...804...27A} {804, 27}

\bibitem[\protect\citeauthoryear{{Assef} et~al.,}{{Assef}
  et~al.}{2016}]{Assef16}
{Assef} R.~J.,  et~al., 2016, \mn@doi [\apj] {10.3847/0004-637X/819/2/111},
  \href {https://ui.adsabs.harvard.edu/abs/2016ApJ...819..111A} {819, 111}

\bibitem[\protect\citeauthoryear{{Assef} et~al.,}{{Assef}
  et~al.}{2019}]{Assef19}
{Assef} R.~J.,  et~al., 2019, arXiv e-prints, \href
  {https://ui.adsabs.harvard.edu/abs/2019arXiv190504320A} {p. arXiv:1905.04320}

\bibitem[\protect\citeauthoryear{{Ba{\~n}ados} et~al.,}{{Ba{\~n}ados}
  et~al.}{2018}]{Banados18b}
{Ba{\~n}ados} E.,  et~al., 2018, \mn@doi [\apj] {10.3847/2041-8213/aab61e},
  \href {https://ui.adsabs.harvard.edu/abs/2018ApJ...856L..25B} {856, L25}

\bibitem[\protect\citeauthoryear{{Ballo}, {Severgnini}, {Della Ceca},
  {Caccianiga}, {Vignali}, {Carrera}, {Corral}  \& {Mateos}}{{Ballo}
  et~al.}{2014}]{Ballo14}
{Ballo} L.,  {Severgnini} P.,  {Della Ceca} R.,  {Caccianiga} A.,  {Vignali}
  C.,  {Carrera} F.~J.,  {Corral} A.,   {Mateos} S.,  2014, \mn@doi [\mnras]
  {10.1093/mnras/stu1628}, \href
  {https://ui.adsabs.harvard.edu/abs/2014MNRAS.444.2580B} {444, 2580}

\bibitem[\protect\citeauthoryear{{Balokovi{\'c}} et~al.,}{{Balokovi{\'c}}
  et~al.}{2018}]{Balokovic18}
{Balokovi{\'c}} M.,  et~al., 2018, \mn@doi [\apj] {10.3847/1538-4357/aaa7eb},
  \href {https://ui.adsabs.harvard.edu/abs/2018ApJ...854...42B} {854, 42}

\bibitem[\protect\citeauthoryear{{Banerji}, {McMahon}, {Hewett},
  {Alaghband-Zadeh}, {Gonzalez-Solares}, {Venemans}  \& {Hawthorn}}{{Banerji}
  et~al.}{2012}]{Banerji12}
{Banerji} M.,  {McMahon} R.~G.,  {Hewett} P.~C.,  {Alaghband-Zadeh} S.,
  {Gonzalez-Solares} E.,  {Venemans} B.~P.,   {Hawthorn} M.~J.,  2012, \mn@doi
  [\mnras] {10.1111/j.1365-2966.2012.22099.x}, \href
  {http://adsabs.harvard.edu/abs/2012MNRAS.427.2275B} {427, 2275}

\bibitem[\protect\citeauthoryear{{Banerji}, {McMahon}, {Hewett},
  {Gonzalez-Solares}  \& {Koposov}}{{Banerji} et~al.}{2013}]{Banerji13}
{Banerji} M.,  {McMahon} R.~G.,  {Hewett} P.~C.,  {Gonzalez-Solares} E.,
  {Koposov} S.~E.,  2013, \mn@doi [\mnras] {10.1093/mnrasl/sls023}, \href
  {https://ui.adsabs.harvard.edu/abs/2013MNRAS.429L..55B} {429, L55}

\bibitem[\protect\citeauthoryear{{Banerji}, {Fabian}  \& {McMahon}}{{Banerji}
  et~al.}{2014}]{Banerji14}
{Banerji} M.,  {Fabian} A.~C.,   {McMahon} R.~G.,  2014, \mn@doi [\mnras]
  {10.1093/mnrasl/slt178}, \href
  {http://adsabs.harvard.edu/abs/2014MNRAS.439L..51B} {439, L51}

\bibitem[\protect\citeauthoryear{{Banerji}, {Alaghband-Zadeh}, {Hewett}  \&
  {McMahon}}{{Banerji} et~al.}{2015}]{Banerji15b}
{Banerji} M.,  {Alaghband-Zadeh} S.,  {Hewett} P.~C.,   {McMahon} R.~G.,  2015,
  \mn@doi [\mnras] {10.1093/mnras/stu2649}, \href
  {http://adsabs.harvard.edu/abs/2015MNRAS.447.3368B} {447, 3368}

\bibitem[\protect\citeauthoryear{{Banerji}, {Jones}, {Wagg}, {Carilli},
  {Bisbas}  \& {Hewett}}{{Banerji} et~al.}{2018}]{Banerji18}
{Banerji} M.,  {Jones} G.~C.,  {Wagg} J.,  {Carilli} C.~L.,  {Bisbas} T.~G.,
  {Hewett} P.~C.,  2018, \mn@doi [\mnras] {10.1093/mnras/sty1443}, \href
  {https://ui.adsabs.harvard.edu/abs/2018MNRAS.479.1154B} {479, 1154}

\bibitem[\protect\citeauthoryear{{B{\"a}r} et~al.,}{{B{\"a}r}
  et~al.}{2019}]{Bar19}
{B{\"a}r} R.~E.,  et~al., 2019, \mn@doi [\mnras] {10.1093/mnras/stz2309}, \href
  {https://ui.adsabs.harvard.edu/abs/2019MNRAS.489.3073B} {489, 3073}

\bibitem[\protect\citeauthoryear{{Bischetti} et~al.,}{{Bischetti}
  et~al.}{2017}]{Bischetti17}
{Bischetti} M.,  et~al., 2017, \mn@doi [\aap] {10.1051/0004-6361/201629301},
  \href {http://adsabs.harvard.edu/abs/2017A%26A...598A.122B} {598, A122}

\bibitem[\protect\citeauthoryear{{Bower}, {Benson}, {Malbon}, {Helly}, {Frenk},
  {Baugh}, {Cole}  \& {Lacey}}{{Bower} et~al.}{2006}]{Bower06}
{Bower} R.~G.,  {Benson} A.~J.,  {Malbon} R.,  {Helly} J.~C.,  {Frenk} C.~S.,
  {Baugh} C.~M.,  {Cole} S.,   {Lacey} C.~G.,  2006, \mn@doi [\mnras]
  {10.1111/j.1365-2966.2006.10519.x}, \href
  {https://ui.adsabs.harvard.edu/abs/2006MNRAS.370..645B} {370, 645}

\bibitem[\protect\citeauthoryear{{Bridge} et~al.,}{{Bridge}
  et~al.}{2013}]{Bridge13}
{Bridge} C.~R.,  et~al., 2013, \mn@doi [\apj] {10.1088/0004-637X/769/2/91},
  \href {https://ui.adsabs.harvard.edu/abs/2013ApJ...769...91B} {769, 91}

\bibitem[\protect\citeauthoryear{{Brusa} et~al.,}{{Brusa}
  et~al.}{2010}]{Brusa10}
{Brusa} M.,  et~al., 2010, \mn@doi [\apj] {10.1088/0004-637X/716/1/348}, \href
  {https://ui.adsabs.harvard.edu/abs/2010ApJ...716..348B} {716, 348}

\bibitem[\protect\citeauthoryear{{Brusa} et~al.,}{{Brusa}
  et~al.}{2015a}]{Brusa15a}
{Brusa} M.,  et~al., 2015a, \mn@doi [\mnras] {10.1093/mnras/stu2117}, \href
  {https://ui.adsabs.harvard.edu/abs/2015MNRAS.446.2394B} {446, 2394}

\bibitem[\protect\citeauthoryear{{Brusa} et~al.,}{{Brusa}
  et~al.}{2015b}]{Brusa15b}
{Brusa} M.,  et~al., 2015b, \mn@doi [\aap] {10.1051/0004-6361/201425491}, \href
  {https://ui.adsabs.harvard.edu/abs/2015A&A...578A..11B} {578, A11}

\bibitem[\protect\citeauthoryear{{Brusa} et~al.,}{{Brusa}
  et~al.}{2016}]{Brusa16}
{Brusa} M.,  et~al., 2016, \mn@doi [\aap] {10.1051/0004-6361/201527900}, \href
  {https://ui.adsabs.harvard.edu/abs/2016A&A...588A..58B} {588, A58}

\bibitem[\protect\citeauthoryear{{Burlon}, {Ajello}, {Greiner}, {Comastri},
  {Merloni}  \& {Gehrels}}{{Burlon} et~al.}{2011}]{Burlon11}
{Burlon} D.,  {Ajello} M.,  {Greiner} J.,  {Comastri} A.,  {Merloni} A.,
  {Gehrels} N.,  2011, \mn@doi [\apj] {10.1088/0004-637X/728/1/58}, \href
  {http://adsabs.harvard.edu/abs/2011ApJ...728...58B} {728, 58}

\bibitem[\protect\citeauthoryear{{Burtscher} et~al.,}{{Burtscher}
  et~al.}{2016}]{Burtscher16}
{Burtscher} L.,  et~al., 2016, \mn@doi [\aap] {10.1051/0004-6361/201527575},
  \href {https://ui.adsabs.harvard.edu/abs/2016A&A...586A..28B} {586, A28}

\bibitem[\protect\citeauthoryear{{Carrera}, {Fern{\'a}ndez-Manteca}  \&
  {Mateos}}{{Carrera} et~al.}{2017}]{Carrera17}
{Carrera} F.~J.,  {Fern{\'a}ndez-Manteca} P.,   {Mateos} S.,  2017, in
  Highlights on Spanish Astrophysics IX. pp 120--124

\bibitem[\protect\citeauthoryear{{Chen} et~al.,}{{Chen} et~al.}{2017}]{Chen17}
{Chen} C.-T.~J.,  et~al., 2017, \mn@doi [\apj] {10.3847/1538-4357/837/2/145},
  \href {https://ui.adsabs.harvard.edu/abs/2017ApJ...837..145C} {837, 145}

\bibitem[\protect\citeauthoryear{{Coatman}, {Hewett}, {Banerji}, {Richards},
  {Hennawi}  \& {Prochaska}}{{Coatman} et~al.}{2019}]{Coatman19}
{Coatman} L.,  {Hewett} P.~C.,  {Banerji} M.,  {Richards} G.~T.,  {Hennawi}
  J.~F.,   {Prochaska} J.~X.,  2019, \mn@doi [\mnras] {10.1093/mnras/stz1167},
  \href {https://ui.adsabs.harvard.edu/abs/2019MNRAS.486.5335C} {486, 5335}

\bibitem[\protect\citeauthoryear{{Costa}, {Rosdahl}, {Sijacki}  \&
  {Haehnelt}}{{Costa} et~al.}{2018}]{Costa18b}
{Costa} T.,  {Rosdahl} J.,  {Sijacki} D.,   {Haehnelt} M.~G.,  2018, \mn@doi
  [\mnras] {10.1093/mnras/sty1514}, \href
  {https://ui.adsabs.harvard.edu/abs/2018MNRAS.479.2079C} {479, 2079}

\bibitem[\protect\citeauthoryear{{Davies} et~al.,}{{Davies}
  et~al.}{2015}]{Davies15}
{Davies} R.~I.,  et~al., 2015, \mn@doi [\apj] {10.1088/0004-637X/806/1/127},
  \href {http://adsabs.harvard.edu/abs/2015ApJ...806..127D} {806, 127}

\bibitem[\protect\citeauthoryear{{Debuhr}, {Quataert}  \& {Ma}}{{Debuhr}
  et~al.}{2011}]{Debuhr11}
{Debuhr} J.,  {Quataert} E.,   {Ma} C.-P.,  2011, \mn@doi [\mnras]
  {10.1111/j.1365-2966.2010.17992.x}, \href
  {https://ui.adsabs.harvard.edu/abs/2011MNRAS.412.1341D} {412, 1341}

\bibitem[\protect\citeauthoryear{{D{\'{\i}}az-Santos}
  et~al.,}{{D{\'{\i}}az-Santos} et~al.}{2016}]{DiazSantos16}
{D{\'{\i}}az-Santos} T.,  et~al., 2016, \mn@doi [\apjl]
  {10.3847/2041-8205/816/1/L6}, \href
  {https://ui.adsabs.harvard.edu/abs/2016ApJ...816L...6D} {816, L6}

\bibitem[\protect\citeauthoryear{{D{\'\i}az-Santos} et~al.,}{{D{\'\i}az-Santos}
  et~al.}{2018}]{DiazSantos18}
{D{\'\i}az-Santos} T.,  et~al., 2018, \mn@doi [Science]
  {10.1126/science.aap7605}, \href
  {https://ui.adsabs.harvard.edu/abs/2018Sci...362.1034D} {362, 1034}

\bibitem[\protect\citeauthoryear{{Edge}, {Sutherland}, {Kuijken}, {Driver},
  {McMahon}, {Eales}  \& {Emerson}}{{Edge} et~al.}{2013}]{Edge13}
{Edge} A.,  {Sutherland} W.,  {Kuijken} K.,  {Driver} S.,  {McMahon} R.,
  {Eales} S.,   {Emerson} J.~P.,  2013, The Messenger, \href
  {http://adsabs.harvard.edu/abs/2013Msngr.154...32E} {154, 32}

\bibitem[\protect\citeauthoryear{{Eisenhardt} et~al.,}{{Eisenhardt}
  et~al.}{2012}]{Eisenhardt12}
{Eisenhardt} P.~R.~M.,  et~al., 2012, \mn@doi [\apj]
  {10.1088/0004-637X/755/2/173}, \href
  {http://adsabs.harvard.edu/abs/2012ApJ...755..173E} {755, 173}

\bibitem[\protect\citeauthoryear{{Elvis} et~al.,}{{Elvis}
  et~al.}{1994}]{Elvis94}
{Elvis} M.,  et~al., 1994, \mn@doi [\apjs] {10.1086/192093}, \href
  {https://ui.adsabs.harvard.edu/abs/1994ApJS...95....1E} {95, 1}

\bibitem[\protect\citeauthoryear{{Fabian}}{{Fabian}}{2012}]{Fabian12}
{Fabian} A.~C.,  2012, \mn@doi [\araa] {10.1146/annurev-astro-081811-125521},
  \href {https://ui.adsabs.harvard.edu/abs/2012ARA&A..50..455F} {50, 455}

\bibitem[\protect\citeauthoryear{{Fabian}, {Celotti}  \& {Erlund}}{{Fabian}
  et~al.}{2006}]{Fabian06b}
{Fabian} A.~C.,  {Celotti} A.,   {Erlund} M.~C.,  2006, \mn@doi [\mnras]
  {10.1111/j.1745-3933.2006.00234.x}, \href
  {https://ui.adsabs.harvard.edu/abs/2006MNRAS.373L..16F} {373, L16}

\bibitem[\protect\citeauthoryear{{Fabian}, {Vasudevan}  \& {Gandhi}}{{Fabian}
  et~al.}{2008}]{Fabian08}
{Fabian} A.~C.,  {Vasudevan} R.~V.,   {Gandhi} P.,  2008, \mn@doi [\mnras]
  {10.1111/j.1745-3933.2008.00430.x}, \href
  {http://adsabs.harvard.edu/abs/2008MNRAS.385L..43F} {385, L43}

\bibitem[\protect\citeauthoryear{{Fabian}, {Vasudevan}, {Mushotzky}, {Winter}
  \& {Reynolds}}{{Fabian} et~al.}{2009}]{Fabian09}
{Fabian} A.~C.,  {Vasudevan} R.~V.,  {Mushotzky} R.~F.,  {Winter} L.~M.,
  {Reynolds} C.~S.,  2009, \mn@doi [\mnras] {10.1111/j.1745-3933.2009.00617.x},
  \href {http://adsabs.harvard.edu/abs/2009MNRAS.394L..89F} {394, L89}

\bibitem[\protect\citeauthoryear{{Fabian}, {Lohfink}, {Kara}, {Parker},
  {Vasudevan}  \& {Reynolds}}{{Fabian} et~al.}{2015}]{Fabian15}
{Fabian} A.~C.,  {Lohfink} A.,  {Kara} E.,  {Parker} M.~L.,  {Vasudevan} R.,
  {Reynolds} C.~S.,  2015, \mn@doi [\mnras] {10.1093/mnras/stv1218}, \href
  {https://ui.adsabs.harvard.edu/abs/2015MNRAS.451.4375F} {451, 4375}

\bibitem[\protect\citeauthoryear{{Fan} et~al.,}{{Fan} et~al.}{2016}]{Fan16a}
{Fan} L.,  et~al., 2016, \mn@doi [\apj] {10.3847/2041-8205/822/2/L32}, \href
  {https://ui.adsabs.harvard.edu/abs/2016ApJ...822L..32F} {822, L32}

\bibitem[\protect\citeauthoryear{{Ferrarese} \& {Merritt}}{{Ferrarese} \&
  {Merritt}}{2000}]{Ferrarese00}
{Ferrarese} L.,  {Merritt} D.,  2000, \mn@doi [\apjl] {10.1086/312838}, \href
  {https://ui.adsabs.harvard.edu/abs/2000ApJ...539L...9F} {539, L9}

\bibitem[\protect\citeauthoryear{{Fiore} et~al.,}{{Fiore}
  et~al.}{2009}]{Fiore09}
{Fiore} F.,  et~al., 2009, \mn@doi [\apj] {10.1088/0004-637X/693/1/447}, \href
  {http://adsabs.harvard.edu/abs/2009ApJ...693..447F} {693, 447}

\bibitem[\protect\citeauthoryear{{Gandhi}, {Horst}, {Smette}, {H{\"o}nig},
  {Comastri}, {Gilli}, {Vignali}  \& {Duschl}}{{Gandhi}
  et~al.}{2009}]{Gandhi09}
{Gandhi} P.,  {Horst} H.,  {Smette} A.,  {H{\"o}nig} S.,  {Comastri} A.,
  {Gilli} R.,  {Vignali} C.,   {Duschl} W.,  2009, \mn@doi [\aap]
  {10.1051/0004-6361/200811368}, \href
  {http://adsabs.harvard.edu/abs/2009A%26A...502..457G} {502, 457}

\bibitem[\protect\citeauthoryear{{Gebhardt} et~al.,}{{Gebhardt}
  et~al.}{2000}]{Gebhardt00}
{Gebhardt} K.,  et~al., 2000, \mn@doi [\apjl] {10.1086/312840}, \href
  {https://ui.adsabs.harvard.edu/abs/2000ApJ...539L..13G} {539, L13}

\bibitem[\protect\citeauthoryear{{Glikman}}{{Glikman}}{2017}]{Glikman17b}
{Glikman} E.,  2017, \mn@doi [Research Notes of the American Astronomical
  Society] {10.3847/2515-5172/aaa0c0}, \href
  {https://ui.adsabs.harvard.edu/abs/2017RNAAS...1a..48G} {1, 48}

\bibitem[\protect\citeauthoryear{{Glikman}, {Helfand}, {White}, {Becker},
  {Gregg}  \& {Lacy}}{{Glikman} et~al.}{2007}]{Glikman07}
{Glikman} E.,  {Helfand} D.~J.,  {White} R.~L.,  {Becker} R.~H.,  {Gregg}
  M.~D.,   {Lacy} M.,  2007, \mn@doi [\apj] {10.1086/521073}, \href
  {https://ui.adsabs.harvard.edu/abs/2007ApJ...667..673G} {667, 673}

\bibitem[\protect\citeauthoryear{{Glikman} et~al.,}{{Glikman}
  et~al.}{2012}]{Glikman12}
{Glikman} E.,  et~al., 2012, \mn@doi [\apj] {10.1088/0004-637X/757/1/51}, \href
  {https://ui.adsabs.harvard.edu/abs/2012ApJ...757...51G} {757, 51}

\bibitem[\protect\citeauthoryear{{Glikman}, {Simmons}, {Mailly}, {Schawinski},
  {Urry}  \& {Lacy}}{{Glikman} et~al.}{2015}]{Glikman15}
{Glikman} E.,  {Simmons} B.,  {Mailly} M.,  {Schawinski} K.,  {Urry} C.~M.,
  {Lacy} M.,  2015, \mn@doi [\apj] {10.1088/0004-637X/806/2/218}, \href
  {https://ui.adsabs.harvard.edu/abs/2015ApJ...806..218G} {806, 218}

\bibitem[\protect\citeauthoryear{{Glikman}, {LaMassa}, {Piconcelli}, {Urry}  \&
  {Lacy}}{{Glikman} et~al.}{2017}]{Glikman17a}
{Glikman} E.,  {LaMassa} S.,  {Piconcelli} E.,  {Urry} M.,   {Lacy} M.,  2017,
  \mn@doi [\apj] {10.3847/1538-4357/aa88ac}, \href
  {https://ui.adsabs.harvard.edu/abs/2017ApJ...847..116G} {847, 116}

\bibitem[\protect\citeauthoryear{{Goulding} et~al.,}{{Goulding}
  et~al.}{2018}]{Goulding18}
{Goulding} A.~D.,  et~al., 2018, \mn@doi [\apj] {10.3847/1538-4357/aab040},
  \href {https://ui.adsabs.harvard.edu/abs/2018ApJ...856....4G} {856, 4}

\bibitem[\protect\citeauthoryear{{G{\"u}ver} \& {{\"O}zel}}{{G{\"u}ver} \&
  {{\"O}zel}}{2009}]{Guver09}
{G{\"u}ver} T.,  {{\"O}zel} F.,  2009, \mn@doi [\mnras]
  {10.1111/j.1365-2966.2009.15598.x}, \href
  {https://ui.adsabs.harvard.edu/abs/2009MNRAS.400.2050G} {400, 2050}

\bibitem[\protect\citeauthoryear{{Hainline}, {Hickox}, {Carroll}, {Myers},
  {DiPompeo}  \& {Trouille}}{{Hainline} et~al.}{2014}]{Hainline14}
{Hainline} K.~N.,  {Hickox} R.~C.,  {Carroll} C.~M.,  {Myers} A.~D.,
  {DiPompeo} M.~A.,   {Trouille} L.,  2014, \mn@doi [\apj]
  {10.1088/0004-637X/795/2/124}, \href
  {https://ui.adsabs.harvard.edu/abs/2014ApJ...795..124H} {795, 124}

\bibitem[\protect\citeauthoryear{{Hamann} et~al.,}{{Hamann}
  et~al.}{2017}]{Hamann17}
{Hamann} F.,  et~al., 2017, \mn@doi [\mnras] {10.1093/mnras/stw2387}, \href
  {http://adsabs.harvard.edu/abs/2017MNRAS.464.3431H} {464, 3431}

\bibitem[\protect\citeauthoryear{{Hickox} \& {Alexander}}{{Hickox} \&
  {Alexander}}{2018}]{Hickox18}
{Hickox} R.~C.,  {Alexander} D.~M.,  2018, \mn@doi [\araa]
  {10.1146/annurev-astro-081817-051803}, \href
  {https://ui.adsabs.harvard.edu/abs/2018ARA&A..56..625H} {56, 625}

\bibitem[\protect\citeauthoryear{{H{\"o}nig}}{{H{\"o}nig}}{2019}]{Honig19}
{H{\"o}nig} S.~F.,  2019, \mn@doi [\apj] {10.3847/1538-4357/ab4591}, \href
  {https://ui.adsabs.harvard.edu/abs/2019ApJ...884..171H} {884, 171}

\bibitem[\protect\citeauthoryear{{H{\"o}nig} \& {Kishimoto}}{{H{\"o}nig} \&
  {Kishimoto}}{2017}]{Honig17}
{H{\"o}nig} S.~F.,  {Kishimoto} M.,  2017, \mn@doi [\apjl]
  {10.3847/2041-8213/aa6838}, \href
  {https://ui.adsabs.harvard.edu/abs/2017ApJ...838L..20H} {838, L20}

\bibitem[\protect\citeauthoryear{{Hopkins}, {Hernquist}, {Cox}  \& {Kere{\v
  s}}}{{Hopkins} et~al.}{2008}]{Hopkins08}
{Hopkins} P.~F.,  {Hernquist} L.,  {Cox} T.~J.,   {Kere{\v s}} D.,  2008,
  \mn@doi [\apjs] {10.1086/524362}, \href
  {http://adsabs.harvard.edu/abs/2008ApJS..175..356H} {175, 356}

\bibitem[\protect\citeauthoryear{{Ichikawa}, {Ueda}, {Terashima}, {Oyabu},
  {Gandhi}, {Matsuta}  \& {Nakagawa}}{{Ichikawa} et~al.}{2012}]{Ichikawa12}
{Ichikawa} K.,  {Ueda} Y.,  {Terashima} Y.,  {Oyabu} S.,  {Gandhi} P.,
  {Matsuta} K.,   {Nakagawa} T.,  2012, \mn@doi [\apj]
  {10.1088/0004-637X/754/1/45}, \href
  {http://adsabs.harvard.edu/abs/2012ApJ...754...45I} {754, 45}

\bibitem[\protect\citeauthoryear{{Ishibashi} \& {Fabian}}{{Ishibashi} \&
  {Fabian}}{2015}]{Ishibashi15}
{Ishibashi} W.,  {Fabian} A.~C.,  2015, \mn@doi [\mnras]
  {10.1093/mnras/stv944}, \href
  {https://ui.adsabs.harvard.edu/abs/2015MNRAS.451...93I} {451, 93}

\bibitem[\protect\citeauthoryear{{Ishibashi} \& {Fabian}}{{Ishibashi} \&
  {Fabian}}{2016}]{Ishibashi16b}
{Ishibashi} W.,  {Fabian} A.~C.,  2016, \mn@doi [\mnras]
  {10.1093/mnras/stw2063}, \href
  {https://ui.adsabs.harvard.edu/abs/2016MNRAS.463.1291I} {463, 1291}

\bibitem[\protect\citeauthoryear{{Ishibashi}, {Fabian}, {Ricci}  \&
  {Celotti}}{{Ishibashi} et~al.}{2018}]{Ishibashi18b}
{Ishibashi} W.,  {Fabian} A.~C.,  {Ricci} C.,   {Celotti} A.,  2018, \mn@doi
  [\mnras] {10.1093/mnras/sty1620}, \href
  {https://ui.adsabs.harvard.edu/abs/2018MNRAS.479.3335I} {479, 3335}

\bibitem[\protect\citeauthoryear{{Jarvis} et~al.,}{{Jarvis}
  et~al.}{2019}]{Jarvis19}
{Jarvis} M.~E.,  et~al., 2019, \mn@doi [\mnras] {10.1093/mnras/stz556}, \href
  {https://ui.adsabs.harvard.edu/abs/2019MNRAS.485.2710J} {485, 2710}

\bibitem[\protect\citeauthoryear{{Just}, {Brandt}, {Shemmer}, {Steffen},
  {Schneider}, {Chartas}  \& {Garmire}}{{Just} et~al.}{2007}]{Just07}
{Just} D.~W.,  {Brandt} W.~N.,  {Shemmer} O.,  {Steffen} A.~T.,  {Schneider}
  D.~P.,  {Chartas} G.,   {Garmire} G.~P.,  2007, \mn@doi [\apj]
  {10.1086/519990}, \href {http://adsabs.harvard.edu/abs/2007ApJ...665.1004J}
  {665, 1004}

\bibitem[\protect\citeauthoryear{{Kakkad} et~al.,}{{Kakkad}
  et~al.}{2016}]{Kakkad16}
{Kakkad} D.,  et~al., 2016, \mn@doi [\aap] {10.1051/0004-6361/201527968}, \href
  {http://adsabs.harvard.edu/abs/2016A%26A...592A.148K} {592, A148}

\bibitem[\protect\citeauthoryear{{Kalberla}, {Burton}, {Hartmann}, {Arnal},
  {Bajaja}, {Morras}  \& {P{\"o}ppel}}{{Kalberla} et~al.}{2005}]{Kalberla05}
{Kalberla} P.~M.~W.,  {Burton} W.~B.,  {Hartmann} D.,  {Arnal} E.~M.,  {Bajaja}
  E.,  {Morras} R.,   {P{\"o}ppel} W.~G.~L.,  2005, \mn@doi [\aap]
  {10.1051/0004-6361:20041864}, \href
  {http://adsabs.harvard.edu/abs/2005A%26A...440..775K} {440, 775}

\bibitem[\protect\citeauthoryear{{Klindt}, {Alexander}, {Rosario}, {Lusso}  \&
  {Fotopoulou}}{{Klindt} et~al.}{2019}]{Klindt19}
{Klindt} L.,  {Alexander} D.~M.,  {Rosario} D.~J.,  {Lusso} E.,   {Fotopoulou}
  S.,  2019, \mn@doi [\mnras] {10.1093/mnras/stz1771}, \href
  {https://ui.adsabs.harvard.edu/abs/2019MNRAS.488.3109K} {488, 3109}

\bibitem[\protect\citeauthoryear{{Kraft}, {Burrows}  \& {Nousek}}{{Kraft}
  et~al.}{1991}]{Kraft91}
{Kraft} R.~P.,  {Burrows} D.~N.,   {Nousek} J.~A.,  1991, \mn@doi [\apj]
  {10.1086/170124}, \href {http://adsabs.harvard.edu/abs/1991ApJ...374..344K}
  {374, 344}

\bibitem[\protect\citeauthoryear{{LaMassa} et~al.,}{{LaMassa}
  et~al.}{2016a}]{LaMassa16a}
{LaMassa} S.~M.,  et~al., 2016a, \mn@doi [\apj] {10.3847/0004-637X/817/2/172},
  \href {https://ui.adsabs.harvard.edu/abs/2016ApJ...817..172L} {817, 172}

\bibitem[\protect\citeauthoryear{{LaMassa} et~al.,}{{LaMassa}
  et~al.}{2016b}]{LaMassa16c}
{LaMassa} S.~M.,  et~al., 2016b, \mn@doi [\apj] {10.3847/0004-637X/820/1/70},
  \href {http://adsabs.harvard.edu/abs/2016ApJ...820...70L} {820, 70}

\bibitem[\protect\citeauthoryear{{LaMassa} et~al.,}{{LaMassa}
  et~al.}{2017}]{LaMassa17}
{LaMassa} S.~M.,  et~al., 2017, \mn@doi [\apj] {10.3847/1538-4357/aa87b5},
  \href {http://adsabs.harvard.edu/abs/2017ApJ...847..100L} {847, 100}

\bibitem[\protect\citeauthoryear{{Lansbury} et~al.,}{{Lansbury}
  et~al.}{2014}]{Lansbury14}
{Lansbury} G.~B.,  et~al., 2014, \mn@doi [\apj] {10.1088/0004-637X/785/1/17},
  \href {http://adsabs.harvard.edu/abs/2014ApJ...785...17L} {785, 17}

\bibitem[\protect\citeauthoryear{{Lanzuisi}, {Piconcelli}, {Fiore}, {Feruglio},
  {Vignali}, {Salvato}  \& {Gruppioni}}{{Lanzuisi} et~al.}{2009}]{Lanzuisi09}
{Lanzuisi} G.,  {Piconcelli} E.,  {Fiore} F.,  {Feruglio} C.,  {Vignali} C.,
  {Salvato} M.,   {Gruppioni} C.,  2009, \mn@doi [\aap]
  {10.1051/0004-6361/200811282}, \href
  {http://adsabs.harvard.edu/abs/2009A%26A...498...67L} {498, 67}

\bibitem[\protect\citeauthoryear{{Lawrence} et~al.,}{{Lawrence}
  et~al.}{2007}]{Lawrence07}
{Lawrence} A.,  et~al., 2007, \mn@doi [\mnras]
  {10.1111/j.1365-2966.2007.12040.x}, \href
  {http://adsabs.harvard.edu/abs/2007MNRAS.379.1599L} {379, 1599}

\bibitem[\protect\citeauthoryear{{Levenson}, {Radomski}, {Packham}, {Mason},
  {Schaefer}  \& {Telesco}}{{Levenson} et~al.}{2009}]{Levenson09}
{Levenson} N.~A.,  {Radomski} J.~T.,  {Packham} C.,  {Mason} R.~E.,  {Schaefer}
  J.~J.,   {Telesco} C.~M.,  2009, \mn@doi [\apj]
  {10.1088/0004-637X/703/1/390}, \href
  {https://ui.adsabs.harvard.edu/abs/2009ApJ...703..390L} {703, 390}

\bibitem[\protect\citeauthoryear{{Liu} et~al.,}{{Liu} et~al.}{2018}]{Liu18}
{Liu} T.,  et~al., 2018, \mn@doi [\mnras] {10.1093/mnras/sty1751}, \href
  {https://ui.adsabs.harvard.edu/abs/2018MNRAS.479.5022L} {479, 5022}

\bibitem[\protect\citeauthoryear{{Luo} et~al.,}{{Luo} et~al.}{2014}]{Luo14}
{Luo} B.,  et~al., 2014, \mn@doi [\apj] {10.1088/0004-637X/794/1/70}, \href
  {http://adsabs.harvard.edu/abs/2014ApJ...794...70L} {794, 70}

\bibitem[\protect\citeauthoryear{{Luo} et~al.,}{{Luo} et~al.}{2015}]{Luo15}
{Luo} B.,  et~al., 2015, \mn@doi [\apj] {10.1088/0004-637X/805/2/122}, \href
  {http://adsabs.harvard.edu/abs/2015ApJ...805..122L} {805, 122}

\bibitem[\protect\citeauthoryear{{Lusso} et~al.,}{{Lusso}
  et~al.}{2012}]{Lusso12}
{Lusso} E.,  et~al., 2012, \mn@doi [\mnras] {10.1111/j.1365-2966.2012.21513.x},
  \href {http://adsabs.harvard.edu/abs/2012MNRAS.425..623L} {425, 623}

\bibitem[\protect\citeauthoryear{{Lutz}, {Maiolino}, {Spoon}  \&
  {Moorwood}}{{Lutz} et~al.}{2004}]{Lutz04}
{Lutz} D.,  {Maiolino} R.,  {Spoon} H.~W.~W.,   {Moorwood} A.~F.~M.,  2004,
  \mn@doi [\aap] {10.1051/0004-6361:20035838}, \href
  {http://adsabs.harvard.edu/abs/2004A%26A...418..465L} {418, 465}

\bibitem[\protect\citeauthoryear{{Magorrian} et~al.,}{{Magorrian}
  et~al.}{1998}]{Magorrian98}
{Magorrian} J.,  et~al., 1998, \mn@doi [\aj] {10.1086/300353}, \href
  {https://ui.adsabs.harvard.edu/abs/1998AJ....115.2285M} {115, 2285}

\bibitem[\protect\citeauthoryear{{Maiolino}, {Marconi}, {Salvati}, {Risaliti},
  {Severgnini}, {Oliva}, {La Franca}  \& {Vanzi}}{{Maiolino}
  et~al.}{2001}]{Maiolino01}
{Maiolino} R.,  {Marconi} A.,  {Salvati} M.,  {Risaliti} G.,  {Severgnini} P.,
  {Oliva} E.,  {La Franca} F.,   {Vanzi} L.,  2001, \mn@doi [\aap]
  {10.1051/0004-6361:20000177}, \href
  {https://ui.adsabs.harvard.edu/abs/2001A&A...365...28M} {365, 28}

\bibitem[\protect\citeauthoryear{{Martocchia} et~al.,}{{Martocchia}
  et~al.}{2017}]{Martocchia17}
{Martocchia} S.,  et~al., 2017, \mn@doi [\aap] {10.1051/0004-6361/201731314},
  \href {https://ui.adsabs.harvard.edu/abs/2017A&A...608A..51M} {608, A51}

\bibitem[\protect\citeauthoryear{{Mateos} et~al.,}{{Mateos}
  et~al.}{2015}]{Mateos15}
{Mateos} S.,  et~al., 2015, \mn@doi [\mnras] {10.1093/mnras/stv299}, \href
  {http://adsabs.harvard.edu/abs/2015MNRAS.449.1422M} {449, 1422}

\bibitem[\protect\citeauthoryear{{McMahon}, {Banerji}, {Gonzalez}, {Koposov},
  {Bejar}, {Lodieu}, {Rebolo}  \& {VHS Collaboration}}{{McMahon}
  et~al.}{2013}]{McMahon13}
{McMahon} R.~G.,  {Banerji} M.,  {Gonzalez} E.,  {Koposov} S.~E.,  {Bejar}
  V.~J.,  {Lodieu} N.,  {Rebolo} R.,   {VHS Collaboration} 2013, The Messenger,
  \href {http://adsabs.harvard.edu/abs/2013Msngr.154...35M} {154, 35}

\bibitem[\protect\citeauthoryear{{Merloni} et~al.,}{{Merloni}
  et~al.}{2014}]{Merloni14}
{Merloni} A.,  et~al., 2014, \mn@doi [\mnras] {10.1093/mnras/stt2149}, \href
  {http://adsabs.harvard.edu/abs/2014MNRAS.437.3550M} {437, 3550}

\bibitem[\protect\citeauthoryear{{Murray}, {Quataert}  \& {Thompson}}{{Murray}
  et~al.}{2005}]{Murray05}
{Murray} N.,  {Quataert} E.,   {Thompson} T.~A.,  2005, \mn@doi [\apj]
  {10.1086/426067}, \href
  {https://ui.adsabs.harvard.edu/abs/2005ApJ...618..569M} {618, 569}

\bibitem[\protect\citeauthoryear{{Narayanan} et~al.,}{{Narayanan}
  et~al.}{2010}]{Narayanan10}
{Narayanan} D.,  et~al., 2010, \mn@doi [\mnras]
  {10.1111/j.1365-2966.2010.16997.x}, \href
  {https://ui.adsabs.harvard.edu/abs/2010MNRAS.407.1701N} {407, 1701}

\bibitem[\protect\citeauthoryear{{Netzer}, {Shemmer}, {Maiolino}, {Oliva},
  {Croom}, {Corbett}  \& {di Fabrizio}}{{Netzer} et~al.}{2004}]{Netzer04}
{Netzer} H.,  {Shemmer} O.,  {Maiolino} R.,  {Oliva} E.,  {Croom} S.,
  {Corbett} E.,   {di Fabrizio} L.,  2004, \mn@doi [\apj] {10.1086/423608},
  \href {https://ui.adsabs.harvard.edu/abs/2004ApJ...614..558N} {614, 558}

\bibitem[\protect\citeauthoryear{{Perrotta}, {Hamann}, {Zakamska}, {Alexand
  roff}, {Rupke}  \& {Wylezalek}}{{Perrotta} et~al.}{2019}]{Perrotta19}
{Perrotta} S.,  {Hamann} F.,  {Zakamska} N.~L.,  {Alexand roff} R.~M.,  {Rupke}
  D.,   {Wylezalek} D.,  2019, \mn@doi [\mnras] {10.1093/mnras/stz1993}, \href
  {https://ui.adsabs.harvard.edu/abs/2019MNRAS.488.4126P} {488, 4126}

\bibitem[\protect\citeauthoryear{{Piconcelli} et~al.,}{{Piconcelli}
  et~al.}{2015}]{Piconcelli15}
{Piconcelli} E.,  et~al., 2015, \mn@doi [\aap] {10.1051/0004-6361/201425324},
  \href {https://ui.adsabs.harvard.edu/abs/2015A%26A...574L...9P} {574, L9}

\bibitem[\protect\citeauthoryear{{Raimundo}, {Fabian}, {Bauer}, {Alexand er},
  {Brandt}, {Luo}, {Vasudevan}  \& {Xue}}{{Raimundo} et~al.}{2010}]{Raimundo10}
{Raimundo} S.~I.,  {Fabian} A.~C.,  {Bauer} F.~E.,  {Alexand er} D.~M.,
  {Brandt} W.~N.,  {Luo} B.,  {Vasudevan} R.~V.,   {Xue} Y.~Q.,  2010, \mn@doi
  [\mnras] {10.1111/j.1365-2966.2010.17234.x}, \href
  {https://ui.adsabs.harvard.edu/abs/2010MNRAS.408.1714R} {408, 1714}

\bibitem[\protect\citeauthoryear{{Ricci} et~al.,}{{Ricci}
  et~al.}{2017a}]{Ricci17d}
{Ricci} C.,  et~al., 2017a, \mn@doi [\apjs] {10.3847/1538-4365/aa96ad}, \href
  {https://ui.adsabs.harvard.edu/abs/2017ApJS..233...17R} {233, 17}

\bibitem[\protect\citeauthoryear{{Ricci} et~al.,}{{Ricci}
  et~al.}{2017b}]{Ricci17c}
{Ricci} C.,  et~al., 2017b, \mn@doi [\nat] {10.1038/nature23906}, \href
  {http://adsabs.harvard.edu/abs/2017Natur.549..488R} {549, 488}

\bibitem[\protect\citeauthoryear{{Ricci} et~al.,}{{Ricci}
  et~al.}{2017c}]{Ricci17a}
{Ricci} C.,  et~al., 2017c, \mn@doi [\apj] {10.3847/1538-4357/835/1/105}, \href
  {https://ui.adsabs.harvard.edu/abs/2017ApJ...835..105R} {835, 105}

\bibitem[\protect\citeauthoryear{{Richards} et~al.,}{{Richards}
  et~al.}{2003}]{Richards03}
{Richards} G.~T.,  et~al., 2003, \mn@doi [\aj] {10.1086/377014}, \href
  {https://ui.adsabs.harvard.edu/abs/2003AJ....126.1131R} {126, 1131}

\bibitem[\protect\citeauthoryear{{Richards} et~al.,}{{Richards}
  et~al.}{2006}]{Richards06a}
{Richards} G.~T.,  et~al., 2006, \mn@doi [\aj] {10.1086/503559}, \href
  {https://ui.adsabs.harvard.edu/abs/2006AJ....131.2766R} {131, 2766}

\bibitem[\protect\citeauthoryear{{Rivers}, {Markowitz}  \&
  {Rothschild}}{{Rivers} et~al.}{2013}]{Rivers13}
{Rivers} E.,  {Markowitz} A.,   {Rothschild} R.,  2013, \mn@doi [\apj]
  {10.1088/0004-637X/772/2/114}, \href
  {https://ui.adsabs.harvard.edu/abs/2013ApJ...772..114R} {772, 114}

\bibitem[\protect\citeauthoryear{{Ross} et~al.,}{{Ross} et~al.}{2015}]{Ross15}
{Ross} N.~P.,  et~al., 2015, \mn@doi [\mnras] {10.1093/mnras/stv1710}, \href
  {http://adsabs.harvard.edu/abs/2015MNRAS.453.3932R} {453, 3932}

\bibitem[\protect\citeauthoryear{{Roth}, {Kasen}, {Hopkins}  \&
  {Quataert}}{{Roth} et~al.}{2012}]{Roth12}
{Roth} N.,  {Kasen} D.,  {Hopkins} P.~F.,   {Quataert} E.,  2012, \mn@doi
  [\apj] {10.1088/0004-637X/759/1/36}, \href
  {https://ui.adsabs.harvard.edu/abs/2012ApJ...759...36R} {759, 36}

\bibitem[\protect\citeauthoryear{{Sanders} \& {Mirabel}}{{Sanders} \&
  {Mirabel}}{1996}]{Sanders96}
{Sanders} D.~B.,  {Mirabel} I.~F.,  1996, \mn@doi [\araa]
  {10.1146/annurev.astro.34.1.749}, \href
  {https://ui.adsabs.harvard.edu/abs/1996ARA&A..34..749S} {34, 749}

\bibitem[\protect\citeauthoryear{{Savage} \& {Mathis}}{{Savage} \&
  {Mathis}}{1979}]{Savage79}
{Savage} B.~D.,  {Mathis} J.~S.,  1979, \araa, \href
  {https://ui.adsabs.harvard.edu/abs/1979ARA%26A..17...73S} {17, 73}

\bibitem[\protect\citeauthoryear{{Shen} et~al.,}{{Shen} et~al.}{2011}]{Shen11}
{Shen} Y.,  et~al., 2011, \mn@doi [\apjs] {10.1088/0067-0049/194/2/45}, \href
  {http://adsabs.harvard.edu/abs/2011ApJS..194...45S} {194, 45}

\bibitem[\protect\citeauthoryear{{Shimizu} et~al.,}{{Shimizu}
  et~al.}{2018}]{Shimizu18}
{Shimizu} T.~T.,  et~al., 2018, \mn@doi [\apj] {10.3847/1538-4357/aab09e},
  \href {https://ui.adsabs.harvard.edu/abs/2018ApJ...856..154S} {856, 154}

\bibitem[\protect\citeauthoryear{{Somerville}, {Hopkins}, {Cox}, {Robertson}
  \& {Hernquist}}{{Somerville} et~al.}{2008}]{Somerville08}
{Somerville} R.~S.,  {Hopkins} P.~F.,  {Cox} T.~J.,  {Robertson} B.~E.,
  {Hernquist} L.,  2008, \mn@doi [\mnras] {10.1111/j.1365-2966.2008.13805.x},
  \href {https://ui.adsabs.harvard.edu/abs/2008MNRAS.391..481S} {391, 481}

\bibitem[\protect\citeauthoryear{{Springel}, {Di Matteo}  \&
  {Hernquist}}{{Springel} et~al.}{2005}]{Springel05}
{Springel} V.,  {Di Matteo} T.,   {Hernquist} L.,  2005, \mn@doi [\apjl]
  {10.1086/428772}, \href
  {https://ui.adsabs.harvard.edu/abs/2005ApJ...620L..79S} {620, L79}

\bibitem[\protect\citeauthoryear{{Stern}}{{Stern}}{2015}]{Stern15}
{Stern} D.,  2015, \mn@doi [\apj] {10.1088/0004-637X/807/2/129}, \href
  {http://adsabs.harvard.edu/abs/2015ApJ...807..129S} {807, 129}

\bibitem[\protect\citeauthoryear{{Stern} et~al.,}{{Stern}
  et~al.}{2014}]{Stern14}
{Stern} D.,  et~al., 2014, \mn@doi [\apj] {10.1088/0004-637X/794/2/102}, \href
  {http://adsabs.harvard.edu/abs/2014ApJ...794..102S} {794, 102}

\bibitem[\protect\citeauthoryear{{Temple}, {Banerji}, {Hewett}, {Coatman},
  {Maddox}  \& {Peroux}}{{Temple} et~al.}{2019}]{Temple19}
{Temple} M.~J.,  {Banerji} M.,  {Hewett} P.~C.,  {Coatman} L.,  {Maddox} N.,
  {Peroux} C.,  2019, \mn@doi [\mnras] {10.1093/mnras/stz1420}, \href
  {https://ui.adsabs.harvard.edu/abs/2019MNRAS.487.2594T} {487, 2594}

\bibitem[\protect\citeauthoryear{{Teng} et~al.,}{{Teng} et~al.}{2014}]{Teng14}
{Teng} S.~H.,  et~al., 2014, \mn@doi [\apj] {10.1088/0004-637X/785/1/19}, \href
  {http://adsabs.harvard.edu/abs/2014ApJ...785...19T} {785, 19}

\bibitem[\protect\citeauthoryear{{Thompson}, {Fabian}, {Quataert}  \&
  {Murray}}{{Thompson} et~al.}{2015}]{Thompson15}
{Thompson} T.~A.,  {Fabian} A.~C.,  {Quataert} E.,   {Murray} N.,  2015,
  \mn@doi [\mnras] {10.1093/mnras/stv246}, \href
  {https://ui.adsabs.harvard.edu/abs/2015MNRAS.449..147T} {449, 147}

\bibitem[\protect\citeauthoryear{{Toba}, {Ueda}, {Matsuoka}, {Shidatsu},
  {Nagao}, {Terashima}, {Wang}  \& {Chang}}{{Toba} et~al.}{2019}]{Toba19}
{Toba} Y.,  {Ueda} Y.,  {Matsuoka} K.,  {Shidatsu} M.,  {Nagao} T.,
  {Terashima} Y.,  {Wang} W.-H.,   {Chang} Y.-Y.,  2019, \mn@doi [\mnras]
  {10.1093/mnras/sty3523}, \href
  {https://ui.adsabs.harvard.edu/abs/2019MNRAS.484..196T} {484, 196}

\bibitem[\protect\citeauthoryear{{Tsai} et~al.,}{{Tsai} et~al.}{2015}]{Tsai15}
{Tsai} C.-W.,  et~al., 2015, \mn@doi [\apj] {10.1088/0004-637X/805/2/90}, \href
  {http://adsabs.harvard.edu/abs/2015ApJ...805...90T} {805, 90}

\bibitem[\protect\citeauthoryear{{Tsai} et~al.,}{{Tsai} et~al.}{2018}]{Tsai18}
{Tsai} C.-W.,  et~al., 2018, \mn@doi [\apj] {10.3847/1538-4357/aae698}, \href
  {https://ui.adsabs.harvard.edu/abs/2018ApJ...868...15T} {868, 15}

\bibitem[\protect\citeauthoryear{{Urrutia}, {Lacy}  \& {Becker}}{{Urrutia}
  et~al.}{2008}]{Urrutia08}
{Urrutia} T.,  {Lacy} M.,   {Becker} R.~H.,  2008, \mn@doi [\apj]
  {10.1086/523959}, \href
  {https://ui.adsabs.harvard.edu/abs/2008ApJ...674...80U} {674, 80}

\bibitem[\protect\citeauthoryear{{Vasudevan} \& {Fabian}}{{Vasudevan} \&
  {Fabian}}{2007}]{Vasudevan07}
{Vasudevan} R.~V.,  {Fabian} A.~C.,  2007, \mn@doi [\mnras]
  {10.1111/j.1365-2966.2007.12328.x}, \href
  {http://adsabs.harvard.edu/abs/2007MNRAS.381.1235V} {381, 1235}

\bibitem[\protect\citeauthoryear{{Vasudevan}, {Mushotzky}, {Winter}  \&
  {Fabian}}{{Vasudevan} et~al.}{2009}]{Vasudevan09b}
{Vasudevan} R.~V.,  {Mushotzky} R.~F.,  {Winter} L.~M.,   {Fabian} A.~C.,
  2009, \mn@doi [\mnras] {10.1111/j.1365-2966.2009.15371.x}, \href
  {http://adsabs.harvard.edu/abs/2009MNRAS.399.1553V} {399, 1553}

\bibitem[\protect\citeauthoryear{{Vasudevan}, {Brandt}, {Mushotzky}, {Winter},
  {Baumgartner}, {Shimizu}, {Schneider}  \& {Nousek}}{{Vasudevan}
  et~al.}{2013}]{Vasudevan13}
{Vasudevan} R.~V.,  {Brandt} W.~N.,  {Mushotzky} R.~F.,  {Winter} L.~M.,
  {Baumgartner} W.~H.,  {Shimizu} T.~T.,  {Schneider} D.~P.,   {Nousek} J.,
  2013, \mn@doi [\apj] {10.1088/0004-637X/763/2/111}, \href
  {http://adsabs.harvard.edu/abs/2013ApJ...763..111V} {763, 111}

\bibitem[\protect\citeauthoryear{{Veilleux}, {Cecil}  \&
  {Bland-Hawthorn}}{{Veilleux} et~al.}{2005}]{Veilleux05}
{Veilleux} S.,  {Cecil} G.,   {Bland-Hawthorn} J.,  2005, \mn@doi [\araa]
  {10.1146/annurev.astro.43.072103.150610}, \href
  {https://ui.adsabs.harvard.edu/abs/2005ARA&A..43..769V} {43, 769}

\bibitem[\protect\citeauthoryear{{Vito} et~al.,}{{Vito}
  et~al.}{2018a}]{Vito18b}
{Vito} F.,  et~al., 2018a, \mn@doi [\mnras] {10.1093/mnras/stx3120}, \href
  {https://ui.adsabs.harvard.edu/abs/2018MNRAS.474.4528V} {474, 4528}

\bibitem[\protect\citeauthoryear{{Vito}, {Brandt}, {Luo}, {Shemmer}, {Vignali}
  \& {Gilli}}{{Vito} et~al.}{2018b}]{Vito18c}
{Vito} F.,  {Brandt} W.~N.,  {Luo} B.,  {Shemmer} O.,  {Vignali} C.,   {Gilli}
  R.,  2018b, \mn@doi [\mnras] {10.1093/mnras/sty1765}, \href
  {https://ui.adsabs.harvard.edu/abs/2018MNRAS.479.5335V} {479, 5335}

\bibitem[\protect\citeauthoryear{{Wagner}, {Bicknell}, {Umemura}, {Sutherland }
   \& {Silk}}{{Wagner} et~al.}{2016}]{Wagner16}
{Wagner} A.~Y.,  {Bicknell} G.~V.,  {Umemura} M.,  {Sutherland } R.~S.,
  {Silk} J.,  2016, \mn@doi [Astronomische Nachrichten]
  {10.1002/asna.201512287}, \href
  {https://ui.adsabs.harvard.edu/abs/2016AN....337..167W} {337, 167}

\bibitem[\protect\citeauthoryear{{Weedman}, {Sargsyan}, {Lebouteiller}, {Houck}
   \& {Barry}}{{Weedman} et~al.}{2012}]{Weedman12}
{Weedman} D.,  {Sargsyan} L.,  {Lebouteiller} V.,  {Houck} J.,   {Barry} D.,
  2012, \mn@doi [\apj] {10.1088/0004-637X/761/2/184}, \href
  {http://adsabs.harvard.edu/abs/2012ApJ...761..184W} {761, 184}

\bibitem[\protect\citeauthoryear{{Wethers} et~al.,}{{Wethers}
  et~al.}{2018}]{Wethers18}
{Wethers} C.~F.,  et~al., 2018, \mn@doi [\mnras] {10.1093/mnras/stx3332}, \href
  {https://ui.adsabs.harvard.edu/abs/2018MNRAS.475.3682W} {475, 3682}

\bibitem[\protect\citeauthoryear{{Wu} et~al.,}{{Wu} et~al.}{2011}]{Wu11}
{Wu} J.,  et~al., 2011, \mn@doi [\apj] {10.1088/0004-637X/736/1/28}, \href
  {http://adsabs.harvard.edu/abs/2011ApJ...736...28W} {736, 28}

\bibitem[\protect\citeauthoryear{{Wu} et~al.,}{{Wu} et~al.}{2018}]{Wu18}
{Wu} J.,  et~al., 2018, \mn@doi [\apj] {10.3847/1538-4357/aa9ff3}, \href
  {https://ui.adsabs.harvard.edu/abs/2018ApJ...852...96W} {852, 96}

\bibitem[\protect\citeauthoryear{{Wylezalek} \& {Morganti}}{{Wylezalek} \&
  {Morganti}}{2018}]{Wylezalek18}
{Wylezalek} D.,  {Morganti} R.,  2018, \mn@doi [Nature Astronomy]
  {10.1038/s41550-018-0409-0}, \href
  {https://ui.adsabs.harvard.edu/abs/2018NatAs...2..181W} {2, 181}

\bibitem[\protect\citeauthoryear{{Zakamska} et~al.,}{{Zakamska}
  et~al.}{2016}]{Zakamska16b}
{Zakamska} N.~L.,  et~al., 2016, \mn@doi [\mnras] {10.1093/mnras/stw718}, \href
  {https://ui.adsabs.harvard.edu/abs/2016MNRAS.459.3144Z} {459, 3144}

\bibitem[\protect\citeauthoryear{{Zappacosta} et~al.,}{{Zappacosta}
  et~al.}{2018}]{Zappacosta18b}
{Zappacosta} L.,  et~al., 2018, \mn@doi [\aap] {10.1051/0004-6361/201732557},
  \href {https://ui.adsabs.harvard.edu/abs/2018A&A...618A..28Z} {618, A28}

\makeatother
\end{thebibliography}


\bsp	
\label{lastpage}
\end{document}